# AC Power Cycling Test Setup and Condition Monitoring Tools for SiC-Based Traction Inverters

Masoud Farhadi, *Student Member, IEEE*, Bhanu Teja Vankayalapati, *Student Member, IEEE*, Rahman Sajadi, and Bilal Akin, *Fellow, IEEE*

*Abstract*—AC power cycling tests allow the most realistic reliability assessment by applying close to real stress to the device or module under test to meet functional safety standards, which is highly critical for traction applications. This paper presents a comprehensive guideline and shares critical know-how to develop a 120 KVA AC power cycling test setup for high-power Silicon Carbide (SiC) modules. As of today, traction applications can not generate an early warning signal for drivers to replace critical components on time. For this purpose, the suitable precursors for all dominant failure mechanisms are discussed, and the corresponding condition monitoring tools are proposed to monitor the device aging on power converters. These condition monitoring tools are integrated into the built-in desaturation protection circuit of the electric vehicle (EV) gate driver for low-cost, practical implementation. The on-resistance of all twelve switches is monitored online as a temperature-sensitive electrical parameter (TSEP) to measure the junction temperature of the devices. To avoid heavy processing load in the microcontroller, the out-of-order equivalent time sampling technique is developed for data sampling, which leads to a measurement error of less than 1.5%. In addition, the design considerations regarding common mode noise and aging effect on DESAT protection are investigated, and experimental findings are presented.

## I. Introduction

**P**OWER electronic systems are increasingly used in emerging and mission-profile critical applications such as EVs and aerospace. The data collected from the field reveal that power semiconductors are one of the common sources of failure in these applications [1]–[4]. Therefore, it is crucial to have better insight regarding the failure mechanisms in power switches to meet safety compliances and minimize EV recalls. This is a bigger concern for relatively new Silicon Carbide (SiC) and EV technology due to long-term reliability-related unknowns. The migration from silicon (Si) to SiC is due to the tremendous advantages of SiC devices over Si counterparts. First, SiC has higher band gap width (3.26 eV vs. 1.12 eV) and higher breakdown electric field (2.2 MV/cm vs. 0.3 MV/cm) compared to Si [5]. Therefore, SiC has a higher blocking voltage capability for a given on-resistance ($R_{ds,on}$). These high-voltage power devices can enable a faster-charging rate for EVs, lower number of strings for PV arrays, lower number of components for HVDC etc. Second, SiC has higher thermal conductivity (4.9 W/cm°C vs. 1.5 W/cm°C) which translates to a higher heat transfer rate. Also, the maximum temperature for Si is limited to 150°C, whereas SiC can support 760°C before the device loses its semiconductor properties [6]. Third, SiC has a higher saturation velocity ($2 \times 10^7$ cm/s vs. $1 \times 10^7$ cm/s) compared to Si [5]. This allows SiC to operate at higher current densities and faster switching frequencies. On the other hand, SiC has lower electron mobility (1000 cm²/v.s vs. 1500 cm²/v.s) compared to Si, which results in relatively lower switching frequencies. Overall, the effect of higher saturation velocity is dominant and SiC can operate at higher frequencies resulting in smaller passive components in power electronics systems [7], [8]. Forth, compared to $1.4 \times 10^{10}$ cm⁻³ for Si, the intrinsic carrier concentration of SiC is $8.2 \times 10^9$ cm⁻³. A lower intrinsic carrier concentration results in a lower reverse leakage current, which reduces switching losses [9]. Finally, SiC offers more consistent inversion layer mobility resulting in lower on-resistance variation over temperature. As an example, the on-resistance change over operating temperature can be as high as 44% and 160% for SiC and Si MOSFETs, respectively [10], [11]. Therefore, SiC devices exhibit incremental increase over operating temperature compared to Si counterparts which reduce the cooling requirements and system cost; and is ideal for EV applications.

Although SiC devices offer superior overall performance compared to their Si counterparts in EV applications, there are several reliability-related challenges that need to be addressed. From the package point of view, Young's modulus is roughly three times higher in SiC compared to Si (501 GPa vs. 162 GPa) [12]. Therefore, the stiffness of SiC is higher, and the same thermal cycling creates higher mechanical stress on the package. In addition, the packaging technologies limit the operating temperature of SiC and its specific on-resistance [13]. Therefore, the device market faces growing pressure to reach the theoretical levels of SiC by decreasing the reliability-oriented package margins. From the chip point of view, some of the issues related to the extrinsic defects, Basal plane dislocations, and stacking fault (SF) have been addressed over recent years [14], [15]. However, there are still some challenges regarding gate oxide weakness which are mostly related to the higher density of interface traps and smaller band offset [16].

Power device reliability can be improved through various approaches that have been discussed in the literature. These include the use of hardware redundancy [17], [18], active thermal control [19], [20], and device state of health monitoring [21]–[23]. The basic idea behind hardware redundancy is to provide backup

The authors would like to thank Texas Instrument Automotive Group and Semiconductor Research Corporation (SRC)/Texas Analog Center of Excellence (TxACE) for supporting this research under the Task ID 2810.054. (Corresponding author: Bilal Akin)

Masoud Farhadi, Rahman Sajadi and Bilal Akin are with the Department of Electrical and Computer Engineering, University of Texas at Dallas, Richardson, 75080, USA. (emails: Masoud.Farhadi@utdallas.edu; abdolrahman.sajadi@utdallas.edu; bilal.akin@utdallas.edu).

Bhanu Teja Vankayalapati is with the Texas Instruments, Dallas, TX 75243 USA (e-mail: b-vankayalapati@ti.com).

components or subsystems that can take over in the event of a failure in the primary component or subsystem. This redundancy can improve the overall reliability of the system, as it reduces the likelihood of a single point of failure causing the entire system to fail. Active thermal control is an important technique used to manage the temperature of SiC MOSFETs, which can significantly improve the device's performance and reliability. By using advanced control techniques [24]–[27], active thermal control systems can control losses and ensure that the device operates within a safe temperature range, which can prevent premature failure. Among these techniques, state of health monitoring is considered to be a cost-effective solution that strikes a balance between ensuring system reliability and cost-effectiveness [28]–[30]. The fundamental concept behind state of health monitoring involves monitoring changes in the values of different aging precursors in real-time or at start-ups. These changes in precursor values can be translated into an estimate of the device's state of health, which can be used to generate an early warning signal to prevent unexpected shutdowns of the converter.

To study these weaknesses and related failure modes, it is necessary to conduct accelerated lifetime tests (ALTs) to observe aging mechanisms for various applications such as EVs. The data of ALTs can be used to extrapolate the performance of the device under normal operation, estimate the remaining useful lifetime, and send early warning signal to the EV dashboard. Various ALTs have been used in the literature to observe different degradation mechanisms of SiC MOSFETs using specific stresses. Among different ALTs, High-Temperature Gate Bias (HTGB) and DC power cycling tests have received considerable attention to accelerate gate oxide and package degradations, respectively [31]–[33]. In the HTGB test, a positive or negative constant voltage is applied to the gate-source terminals under high temperatures to accelerate gate oxide degradation. Beyond this, another test with high-frequency gate-source pulses together with high E-field across the drain-source and channel conduction exhibits more realistic operation. In the DC power cycling test, a static thermal cycle is applied to the device by DC current injection at a low voltage within the device's Ohmic region. The thermal cycles during the power cycles causes thermo-mechanical stress and consequently package degradation. In the DC power cycling tests, conduction loss is the only source of heat and there is no switching loss nor hot carrier injection (HCI). Therefore, the DC power cycling test does not represent a real operation.

In [34], a grid-connected active front-end power factor correction (PFC) converter is proposed to consider the gate switching and load current effect with a real mission profile. However, the device is only forward conducted in this test setup. Also, due to the fixed grid frequency, the average temperature is higher than normal test conditions. Therefore, this test mainly triggers package degradation mechanisms. In [35], an IGBT-based ACPC test setup is proposed to circulate current between two back-to-back inverters and eliminate the need for a real load. Even though this test setup is a cost-effective solution, condition monitoring circuits are not developed for online monitoring of aging and thermal indicators. A thermal observer structure is proposed for active thermal control of AC power cycling in [36] where in [37] a methodology for the AC power cycling test is proposed. This test setup is designed to generate low-frequency thermal cycles to decrease the test time.

The paper is organized as follows. In Section II, a concise summary of failure mechanisms and their standard acceleration tests are given to provide essential insight into choosing different operation modes. In Section III, the configuration and control block diagram of the AC power cycling test is discussed in detail. The test setup is fully programmable and can adjust the modulation index, switching frequency, fundamental frequency, DC-link voltage, current amplitude, and power factor to test various operations. The design provides a low-inductance power loop, the capability of carrying high currents (up to 400A), and easy stack-up for assembly. In Section IV, a comprehensive comparison of various aging precursors is presented to identify suitable precursors to monitor in the AC power cycling test. This is especially important for EVs as most of the precursors do not have enough sensitivity and practical implementation capability. Three precursors are selected to monitor the package, body diode, and gate oxide-related degradations with adequate accuracy. Next, in Section V, the corresponding condition monitoring circuits are developed for these selected precursors which are integrated into the DESAT protection circuit to reduce the number of components and ease the implementation. In Section VI, different TSEPs are compared to identify the appropriate indicator of junction temperature. The junction temperature can change rapidly during the converter operation. Therefore, the online signal acquisition and processing for the selected temperature-sensitive parameter is challenging. The out-of-order equivalent time sampling is developed to decrease the signal sampling time. 300 data points are captured out of order for temperature measurement. Since the data are susceptible to noise, an FIR filter is used to filter the raw data. Last but not least, in Section VII, the design considerations for practical issues are discussed in depth.

## II. Degradation Mechanisms and Accelerated Lifetime Tests of SiC MOSFETs

In field applications, different converter operating conditions can initiate various failure mechanisms. Identifying the test setup and adjusting the proper accelerated test conditions are essential to mimic the converter's operating condition. This section presents the dominant failure mechanisms of SiC MOSFETs and various accelerated aging test setups. Figs. 1 (a) and (b) show the SiC MOSFET power module's cross-section, different package-related failure mechanism locations, and their SEM images. SiC MOSFETs failure mechanisms can be broadly classified in three categories: (I) package, (II) gate-oxide, and (III) body diode-related aging mechanisms. The package-related failure mechanisms occur due to the coefficient of thermal expansion (CTE) mismatch among different layers in the device. The power cycling and thermal cycling tests are appropriate tests to accelerate package-related degradation since they apply thermal swings to each layer causing substantial thermo-mechanical stress due to CTE mismatch. The typical locations for package-related failure mechanisms are bond wire flexure, die attachment solder, and surface metallization. CTE mismatch at bond-wire can cause bond-wire lift-off and heel crack, which increases on-resistance [41], [42]. In addition to the thermal cycle, a high drain-current can accelerate bond-wire aging by expanding the internal void in Al bond-wires [43], [44]. CTE mismatch at die attachment solder can increase the junction-to-case thermal resistance and change transient thermal response [45], [46]. Also, SiC has lower CTE compared to Al surface (4.3 ppm/°C vs. 22 ppm/°C) which translates to plastic deformation at high junction temperatures [47], [48].

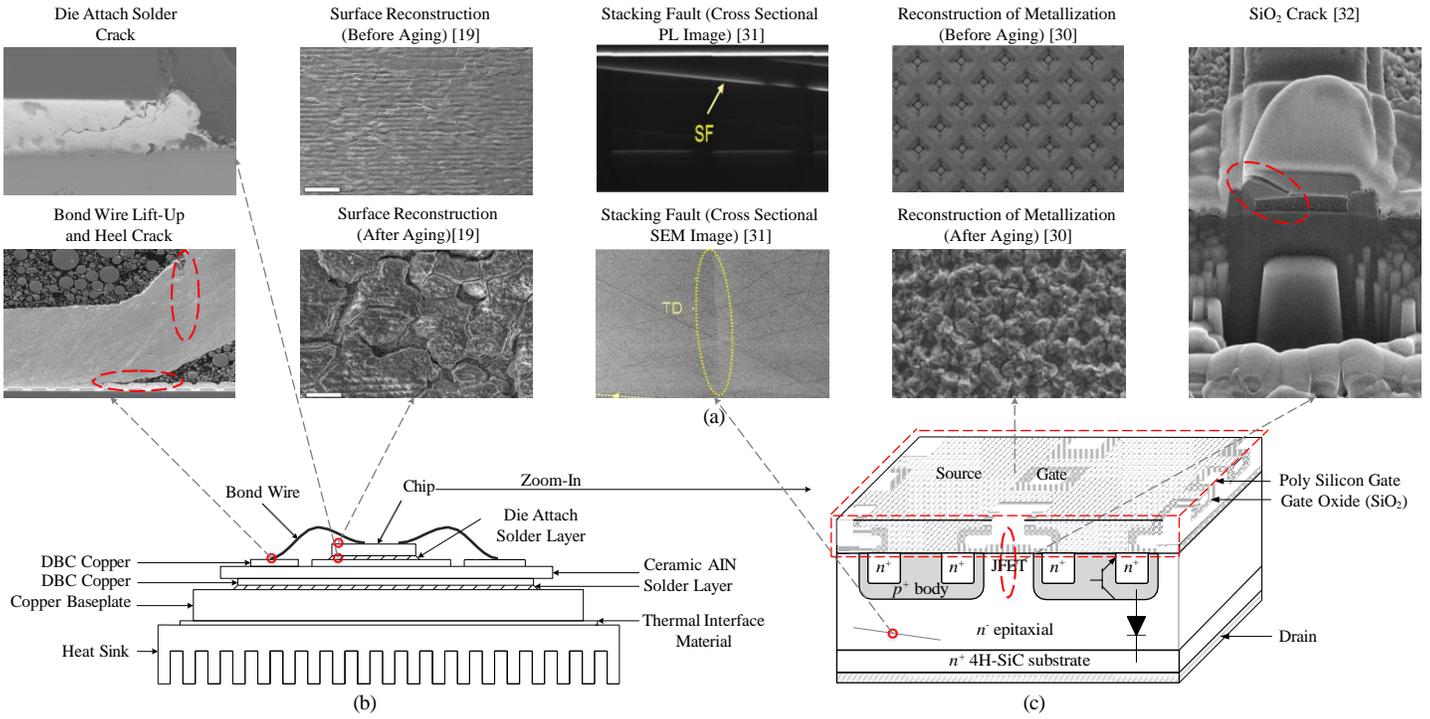

Fig. 1: The cross-section of the SiC MOSFET power module and different failure mechanism locations. (a) Microscopy Images of different failure mechanisms. (b) Cross-section of SiC MOSFET module and locations of common package-related failure mechanisms. (c) Cross-section of SiC MOSFET chip and locations of common gate-oxide and body diode failure mechanisms [35], [38]–[40].

Gate oxide-related failure mechanism is another major reliability challenge for SiC MOSFETs. As SiC has a wider bandgap, it can tolerate a higher electric field which makes SiC/SiO$_2$ interface more susceptible to failure. The oxide electric field can reach 5MV/cm in SiC compared to 3MV/cm in Si counterpart [49]. The oxide thickness can be increased to decrease the oxide electric field. However, the thick gate oxide leads to high threshold voltage and power losses. Also, the interface defect density of SiC/SiO$_2$ is two times higher than Si/SiO$_2$ due to the presence of carbon atoms during the thermal oxidation [50], [51]. During converter operations, the positive or negative gate bias can charge the traps with electrons or holes and make a positive or negative threshold voltage shift known as bias temperature instability (BTI). The positive threshold voltage shift causes higher power loss and the negative threshold voltage shift can increase the cross-talk effect. Additionally, SiC has a three times wider bandgap than Si counterparts and consequently smaller energy band offsets [52]. Therefore, SiC has a higher Fowler-Nordheim tunneling current for the same gate-source voltage. Over time, charged defects line up and make a short-circuit path between gate and source which is known as time-dependent dielectric breakdown (TDDB).

In addition to the high electric field at a high temperature, the channel hot-carrier injection effect can also lead to oxide degradation. During switching transient, MOSFET operates at relatively high drain to source voltage and gate to source voltage, and some carriers have enough energy to pass the oxide barrier [53]. These carriers get trapped in oxide and lead to the hot-carrier injection effect and threshold voltage shift.

The SiC MOSFET's body diode has a lower reverse recovery than its Si MOSFET counterpart. Therefore, using the internal body diode is recommended to reduce cost and increase the power density. On the other hand, the third quadrant conduction can expand defects in the PN junction of SiC MOSFET and degrades its performance [54]. The third quadrant conduction and forward bias expand the basal plane and create a triangular-shaped stacking fault within the drift region [39], [55]–[57]. This expansion increases the threshold voltage and on-resistance, leading to higher power loss. Figs. 1 (a) and (c) show the SiC MOSFET power module chip's cross-section, different gate oxide and body diode-related failure mechanism locations, and their microscopy images.

Several accelerated aging tests are proposed to evaluate the aforementioned failure mechanisms. Fig. 2 shows the accelerated lifetime tests and their aging mechanisms for SiC MOSFETs. The HTGB and high-temperature reverse bias (HTRB) tests are used to gate oxide aging acceleration by applying high temperature and a static voltage to the gate and drain, respectively. In HTGB, drain and source are shorted, and the channel and JFET regions are subject to the electric field. Therefore, TDDB and BTI are expected to contribute to oxide aging [33], [58]. In HTRB, a negative bias is applied to the gate while a positive voltage is applied to the drain-source terminals. These tests apply static stress to the gate oxide. However, applying dynamic tension to mimic the actual operating condition is essential. The high temperature gate switching (HTGS) test uses high-frequency pulses to the gate source for close to the real conditions [59]. This test is more realistic and device under test (DUT) experiences gate-source voltage transient. However, the test time is relatively long, and the channel is not conducted.

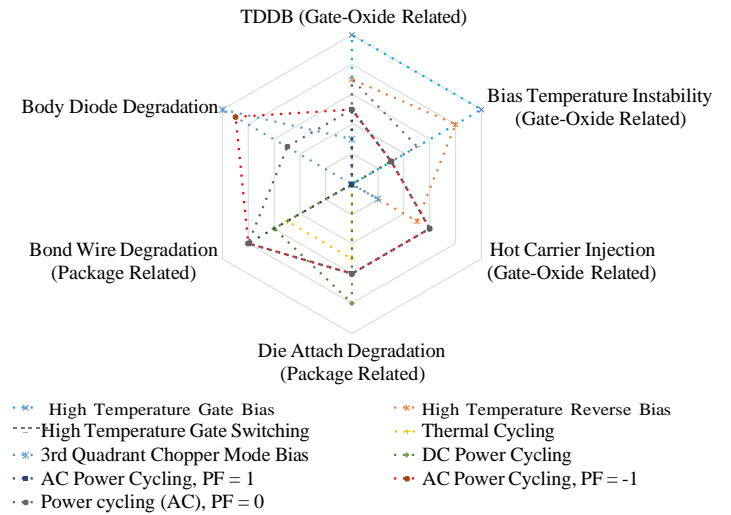

Fig. 2: Accelerated lifetime tests and their aging mechanisms for SiC MOSFETs.

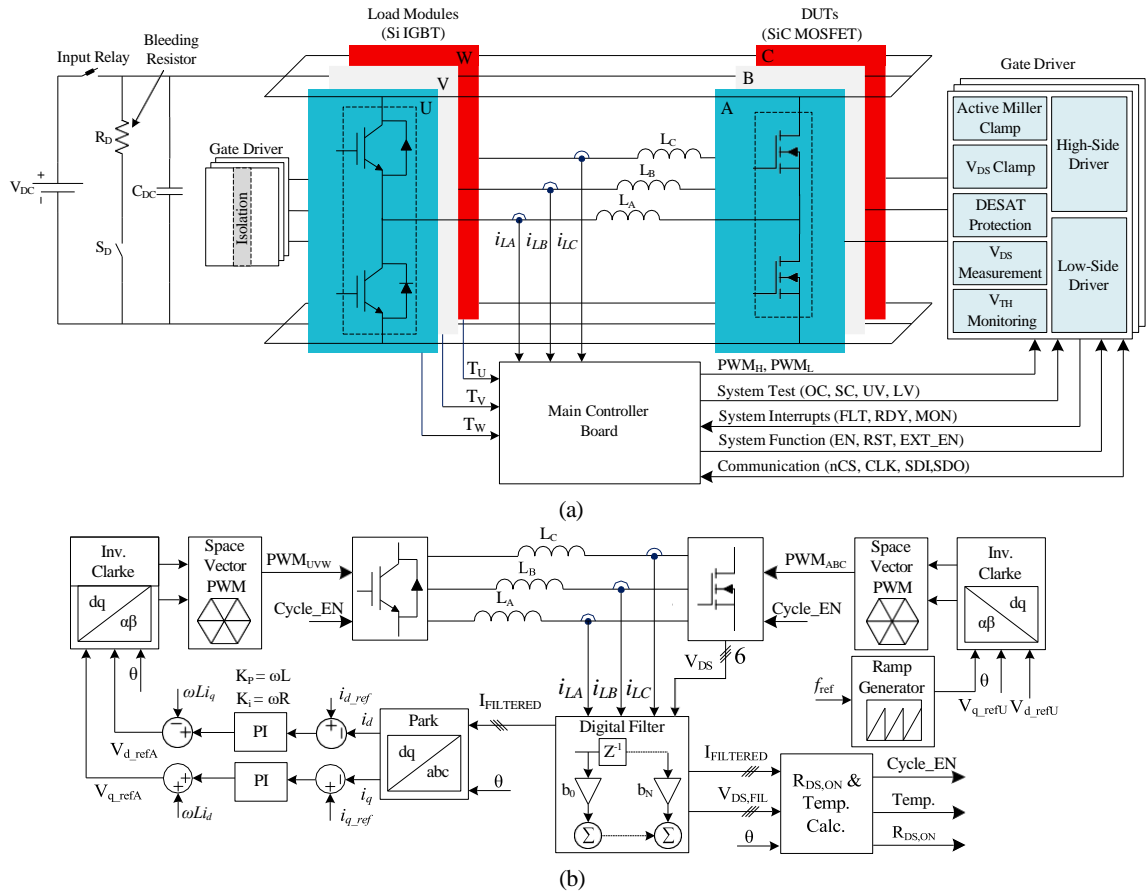

Fig. 3: AC power cycling test configuration and condition monitoring tools for high power SiC MOSFETs. (a) Test setup configuration. (b) Control block diagram.

Thermal cycling and DC power cycling tests accelerate package degradation mechanisms by applying temperature swings to the device. The thermal cycling test can be accomplished through programmable hot plates or ovens with different thermal profiles. This test has no channel conduction, and bond wire self-heating and electro-migration effects are ignored [60]. On the other hand, the DC power cycling test creates thermo-mechanical stress by DC load current injection. In this test, the applied gate-source voltage also causes gate oxide degradation. Depending on test conditions like heat-up time and mean junction temperature, DC power cycling target different parts of the device package [61]. Moreover in this test, the switching transients and high DC bus voltage effects are ignored.

The chopper mode bias (CMB) test accelerates the internal PiN diode degradation by body diode conduction [62], [63]. The device is turned off by negative voltage to ensure the channel is fully off and reverse conducted. The common precursor of body diode degradation is body diode forward voltage ($V_f$). However, $V_f$ can also change due to the package and gate oxide degradation which makes it challenging to extract $V_f$ patterns related to body diode degradation [63].

To address the mentioned issues, the AC power cycling test can create thermal stress under an actual converter operation. The load in this test is AC under DC bus voltage which mimics hot carrier injection and electro-migration effect. Also, both positive and negative gate voltage can be applied to accelerate BTI and TDDB. As shown in Fig. 2, different converter operating conditions can be mimicked by adjusting the power factor (PF), frequency, DC bus voltage, and gate voltages in the AC power cycling test. The configuration and condition monitoring tools for the AC power cycling test are discussed next section in detail.

III. AC POWER CYCLING TEST

A suitable test setup for high-power SiC MOSFET health monitoring is expected to be power efficient while adjusting the losses to accelerate aging mechanisms. To this end, two 3-phase back-to-back inverters are deployed to circulate energy in the AC power cycling test setup. The required heat for thermal cycling is adjusted by fine-tuning the switching and conduction losses at high voltages and loads. Although it is a very high-power setup, the input DC power supply provides only the inverter losses where the bulky part of the power just recycled. Therefore, the overall power requirement to run the test setup is very low. Each phase of the inverters is independently connected to DC bus, and the design consists of three independent full bridge inverters. However, the three-phase configuration allows constant power flow and high DC voltage utilization. Fig. 3 shows the AC power cycling test configuration, control block diagram, and condition monitoring tools for high-power SiC MOSFETs modules. The SiC-based inverter operates as the inverter under test (IUT), and the other IGBT-based inverter operates as an electrical load. The IGBT modules are selected with relatively higher rating than SiC modules as the load inverter needs to run for a longer time. First, setup starts to run and the devices heat up. When the devices in IUT reach the predefined maximum temperature, the setup is turned off, and the cooling system starts to cool the devices down. Each inverter has a separate active liquid cooling system. The cooling system for the load inverter works continuously to decrease IGBTs' degradation rate. However, the cooling system on the test inverter runs during the cooling periods and then shots down when the heat-up cycle starts. Since the cooling down period is the longest part of a thermal swing, high cooling capacity is essential to have a reasonable short thermal cycle. In this setup, two cooling plates are fabricated for 62mm module-based inverters. Each cooling plate can transfer 1.5 KW at a flow rate of 4 liters per minute and a maximum pressure drop of 2.2 psi. As the cooling plates are exposed to high temperatures during the heat-up periods, the cooling system is kept off. Therefore, the copper tube-based cold plates are unsuitable for this application as they use low temperature solder to attach copper tubes to cooling plates. The vacuum brazing technology is better alternative for power cycling test setups and allows metal-to-metal flux-free bonding. A bleeder resistor is connected in parallel to DC bus voltage to discharge the energy stored in DC link capacitors when the setup is turned off. Two laminated busbars with horizontal bussing geometry are fabricated to attach the film capacitors to the power modules. This enables a busbar assembly without bends or standoffs and, hence, a low inductance power loop. Also, the DC link capacitors are placed close to modules to decrease the total power loop inductance and lower peak overshoot voltage. Three 700 $\mu$H, 400 A inductors are fabricated which enable the setup to operate at low current ripple. Fig. 4 shows the real pictures of the AC power cycling test setup, main control board, three-phase load, gate driver board, and cooling system.

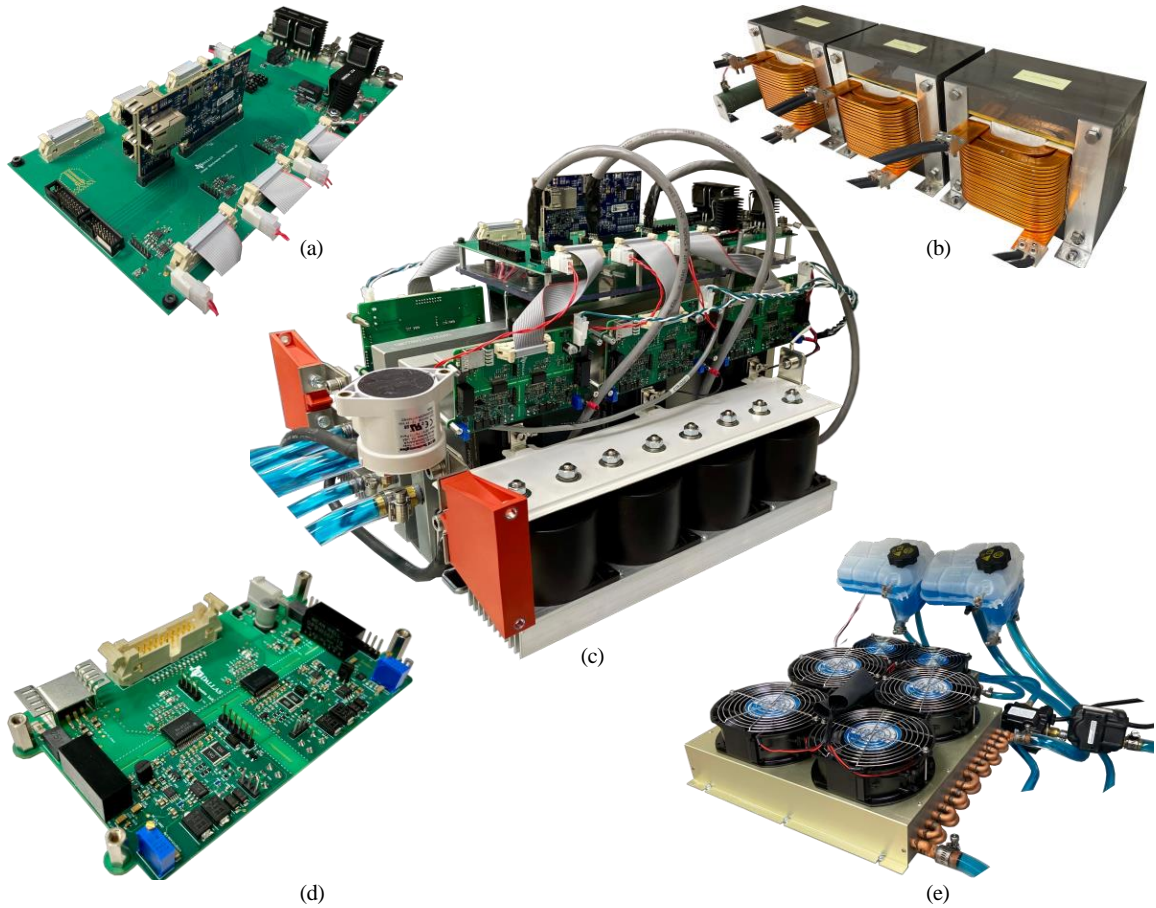

Fig. 4: Picture of the reconfigurable AC power cycling test setup. (a) Main control board. (b) Three-phase load: (3 pc. 700 $\mu$H, 400 A). (c) AC power cycling test setup. (d) Gate driver board with condition monitoring tools. (e) Cooling system.

Gate driver boards are developed using gate driver IC (UCC5870) for each power module and each board has three functions: driving, condition monitoring, and protection. UCC5870 is a high-end gate driver which has peripherals such as integrated ADCs, anti-overlap protection, and active short circuit (ASC). The condition monitoring circuits on these boards measure aging precursors such as threshold voltage, on-resistance, and body diode voltage; and report impending issues in power devices months before they fail. The aging precursors and developed condition monitoring circuits are discussed in the following sections in detail. The gate driver board can apply different turn-on and turn-off voltages to the device to adjust the electric field across gate oxide and, consequently gate oxide degradation rate. An input relay separates the DC power supply from the setup in case of a short circuit. However, the relay is not fast enough to prevent devices from thermal runaway and failure. Therefore, DESAT protection is used as the primary protection for fast short-circuit protection. In addition, secondary overcurrent and overvoltage protection are added through the integrated comparator subsystem (CMPSS) of F28388D C2000 MCU. The drain-source voltage clamp circuit is used to reduce overshoot voltage during device turn-off. Furthermore, an active Miller clamp is used to avoid a false turn-on due to the cross-talk effect. These protections are essential to detect and isolate the faults at the end of lifetime, avoid thermal runaway and device further damage. By fast protection, the aged device can be isolated before serious damage and analyzed to find root cause of failure. This is specially important to take preventative actions in device design process to avoid same failures.

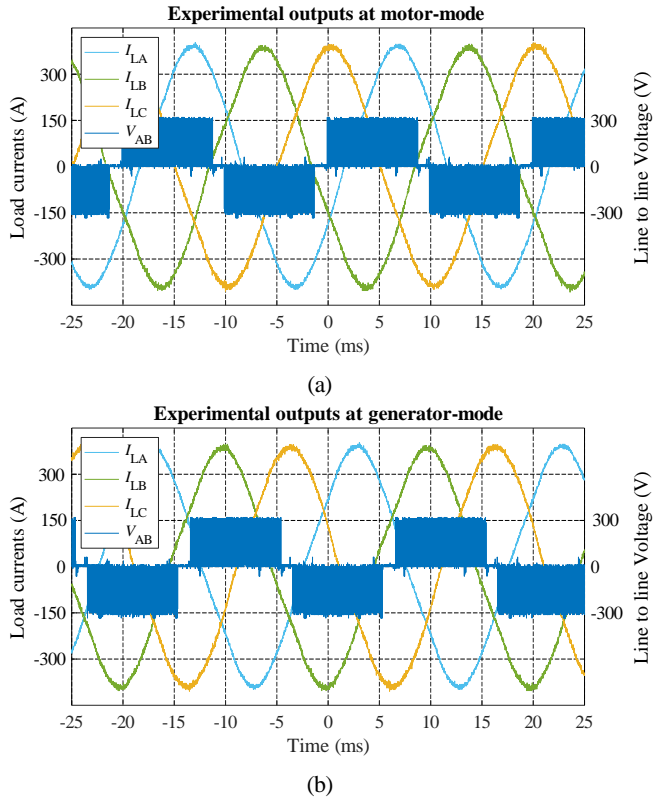

Fig. 5: Experimental waveforms of load currents and line-to-line voltage. (a) Motor-mode ($\theta_V - \theta_i = 0°$). (b) Generator-mode ($\theta_V - \theta_i = 180°$). ($f_{sw}$ = 22 KHz, $f_{fundamental}$ = 50 Hz).

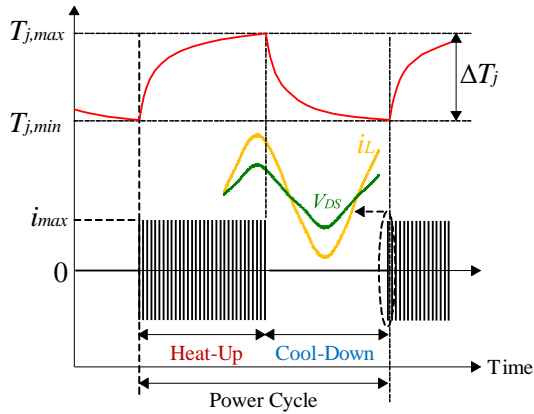

Fig. 6: Illustration of the applied thermal cycle during AC power cycling test.

The gate driver boards and main controller communicate through two address-based SPI interfaces. The inverters are synchronized in the rotating reference frame and controlled by SVM to reduce harmonic content and increase the dc bus voltage utilization. The test inverter is controlled by the open-loop controller to adjust the voltage level and the load inverter is controlled by the closed-loop controller to generate the required current in the test inverter. Fig. 3 (b) shows the control block diagram of test setup. As can be seen here, load currents are measured and applied to a digital filter. Next, the d- and q-axis currents ($i_d$, $i_q$) are calculated through Park transformation in the stationary frame. The error values are calculated based on the measured DQ currents and their reference values. The error values are used as the input of the PI controllers and converted to the reference voltage values for the SV PWM generator. The different power factors can be achieved by adjusting the amplitude and polarity of dq-axis references ($v_{d,test}$, $v_{q,test}$, $i_{d,load}$, and $i_{q,load}$). The modulation index, switching frequency, fundamental frequency, DC-link voltage, current amplitude, and power factor can be adjusted as degrees of freedom. These features make the AC power cycling test setup very flexible to investigate the effects of each factor on the device degradation individually.

The test setup can perform three distinct power cycling techniques. The initial technique involves configuring the setup with a constant $t_{on}$ and $t_{off}$. During this configuration, the test parameters $T_{j,max}$, $T_{j,min}$, and $\Delta T_j$ gradually increase due to the aging effects without the involvement of any compensatory controller. Consequently, the device being tested undergoes substantial stress, leading to early failure. The second power cycling technique regulates $t_{on}$ and $t_{off}$ in a closed-loop system to generate a constant case temperature swing. This configuration provides partial compensation for the aging effects. However, the use of a negative temperature coefficient (NTC) sensor to monitor the case temperature may introduce measurement inaccuracies. The third power cycling technique monitors the junction temperature to produce a constant junction temperature swing (fixed $T_{j,max}$, $T_{j,min}$, and $\Delta T_j$) that can counteract the degradation effects. Temperature measurement is achieved by utilizing the on-resistance as a temperature-sensitive parameter with an accuracy of less than 1.5 %. Therefore, this technique appears to be a more viable option as a power cycling test technique.

Fig. 5 shows the experimental waveforms of load currents and line-to-line voltage in motor- (PF = 1) and generator-mode (PF = -1) modes as examples. In motor mode, package degradation is the dominant failure mechanism. However, in generator mode, the devices work in the third quadrant conduction, and body diode degradation becomes more critical. The setup is run during the heat-up period. When the device temperature reaches to a predefined maximum temperature, the setup is turned off. Fig. 6 illustrates the applied thermal cycle during the AC power cycling test. On-resistance of the device is used as a TSEP to measure and monitor the junction temperature over the thermal cycle, which is discussed in the following sections in detail.

## IV. PRECURSOR SELECTION FOR CONDITION MONITORING DURING AC POWER CYCLING TESTS

It is very critical to monitor the device condition throughout the aging to observe patterns associated with die and package degradations. Different aging precursors have been presented in the literature for condition monitoring of the SiC MOSFETs.

However, of the many precursors proposed, most do not have enough sensitivity or their onboard implementation is not very practical. Selected precursors for the AC power cycle test should be (1) able to cover all dominant failure mechanisms as the AC power cycle test can accelerate most aging mechanisms, (2) highly sensitive to aging related changes, (3) insensitive to load and operating conditions, and (4) easy to measure. However, the aging precursors are typically affected from several aging mechanisms, and decoupling these combined factors is necessary to better understand the individual impact of aging mechanisms on the precursors.

Different aging precursors for degradation mechanisms of SiC MOSFETs are shown in Fig. 7. Threshold voltage $V_{th}$ is a highly sensitive precursor for gate oxide degradation [64]. At the same time, it is insensitive to other degradations, making it a suitable precursor for monitoring in the AC power cycling tests. Online monitoring of the threshold voltage needs to be fast to capture recoverable $V_{th}$ due to high interface state densities [65].

Body diode voltage ($V_{SD}$) and drain leakage current ($I_{DSS}$) are body diode degradation indicators. Between them, $V_{SD}$ has relatively higher sensitivity (up to 0.7 V shift over aging) compared to $I_{DSS}$ (850 nA shift over aging) [66], [67]. Therefore, the body diode voltage is another selected precursor for monitoring in the AC power cycling test. Even though $V_{SD}$ is gate oxide degradation dependent, by applying a negative gate-source voltage to the device gate oxide effect is eliminated. This is because SiC MOSFET conducts only through PiN diode at negative gate-source voltages [68].

On-resistance is reported as an effective precursor for package degradation monitoring. Package degradation events like bond wire lift-up and solder crack can result in a sudden on-resistance increase. At the same time, gate-oxide degradation decreases the overdrive voltage ($V_{GS}$ - $V_{th}$) and can cause a gradual shift in on-resistance [69]. This effect can be compared and compensated by threshold voltage measurement. On-resistance is a suitable precursor for online monitoring, and the same circuit for body-diode voltage monitoring can be used for on-resistance measurement.

Several precursors are proposed based on the device switching transients like turn-off delay time, gate plateau voltage/time, etc [70]–[72]. However, due to the inherent fast switching of SiC MOSFETs, aging monitoring by these precursors needs high sensing resolution and accuracy. The gate leakage current ($I_{GSS}$) is another precursor for gate-oxide degradation monitoring [73]. $I_{GSS}$ is almost constant over the lifetime and increases suddenly at the end of the lifetime. Therefore, it is hard to use it for switch fault diagnosis. The kelvin-source impedance is presented for package degradation monitoring which is only applicable for 4-pin devices with the Kelvin connection [74]. The device's thermal impedance [75], saturation current [76], and miller capacitance [77] are other parameters for aging monitoring. These precursors can hardly be used for online tracking.

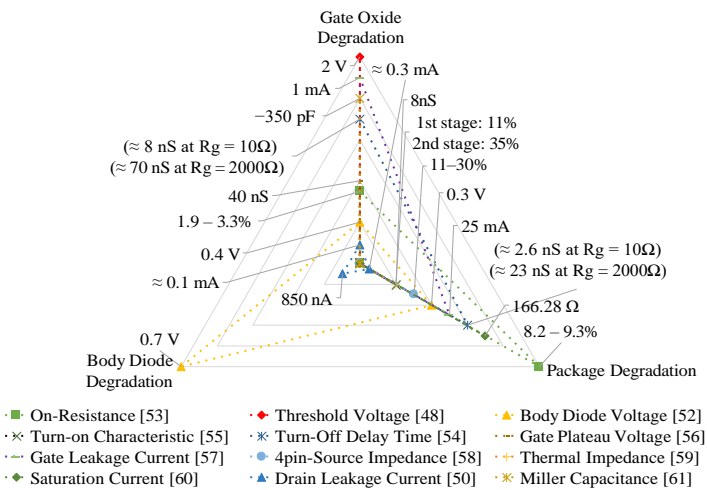

Fig. 7: Aging precursors of different degradation mechanisms for SiC MOSFETs.

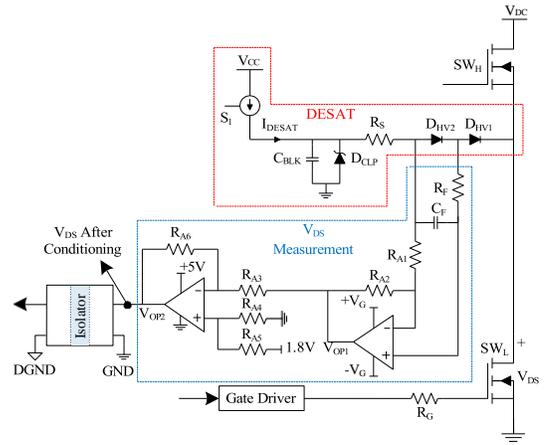

Fig. 8: Drain-source voltage measurement circuit diagram for on-resistance and body-diode voltage monitoring.

In summary, on-resistance, body diode voltage, and threshold voltage are selected to monitor in the AC power cycling test setup. Selected precursors can monitor the package, body diode, and gate oxide-related degradations with sufficient accuracy. Condition monitoring circuits for the selected aging precursors are discussed in the following section.

## V. CONDITION MONITORING CIRCUITS FOR SELECTED PRECURSORS

### A. On-Resistance and Body Diode Voltage Measurement

For on-resistance measurement, the on-state drain-source voltage ($V_{DS,ON}$) is divided by the drain current. The body diode voltage ($V_{SD}$) is measured during the third quadrant operation at the source and drain terminals. Fig. 8 shows the drain-source voltage measurement circuit diagram for on-resistance and body-diode voltage monitoring. The measurement circuit is integrated into DESAT protection circuit to reduce the number of components and cost. Two high voltage fast reverse recovery diodes ($D_{HV1}$, $D_{HV2}$) are connected in series to block DC bus voltage during switch turn-off. These diodes are biased with DESAT protection current during switch turn-on. When the switch ($SW_L$) turns on, the DESAT current flows through the $R_S$, $D_{HV1}$, $D_{HV2}$, and $SW_L$. By applying Kirchhoff's voltage law, the voltage present at the positive terminal of the operational amplifier (OP1) can be derived as follows:

$$V_{OP1,+} = V_{DS} + V_{DHV1} \quad (1)$$

Furthermore, by utilizing the principle of superposition, the voltage present at the negative terminal of the operational ampli-fier (OP1) can be derived as follows:

$$V_{OP1,-} = V_{OP1}(\frac{R_{A1}}{R_{A1} + R_{A2}}) \\ +(V_{DS} + V_{DHV1} + V_{DHV2})(\frac{R_{A2}}{R_{A1} + R_{A2}}) \quad (2)$$

Since op-amp inputs have the same voltages:

$$V_{OP1,+} = V_{OP1,-} \quad (3)$$

Upon substitution of (1) and (2) into (3), the following is obtained:

$$V_{DS} + V_{DHV1} = V_{OP1}(\frac{R_{A1}}{R_{A1} + R_{A2}}) \\ +(V_{DS} + V_{DHV1} + V_{DHV2})(\frac{R_{A2}}{R_{A1} + R_{A2}}) \quad (4)$$

By choosing a same value for $R_{A1}$ and $R_{A2}$ and same part number for $D_{HV1}$ and $D_{HV2}$, the drain-source voltage can be calculated as:

$$V_{OP1} = V_{DS} \quad (5)$$

A high value is recommended for $R_{A1}$ and $R_{A2}$ to decrease mismatch error. Also, there is a small mismatch ($e_D$) between the drop voltages of $D_{HV1}$ and $D_{HV2}$ which can be measured at $I_{DESAT}$. Considering this error, the equation can be rewritten as:

$$V_{OP1} = V_{DS} + e_D \quad (6)$$

In this setup, $I_{DESAT}$ is $1mA$, and $e_D$ ($\approx 0.3 - 1.6mV$) is measured for the DESAT diodes of all switches. A low-pass RC filter ($R_F, C_F$) is used to improve the amplifier's response. In the second stage, the measured voltage is shifted up and scaled properly to meet ADC voltage limitations.

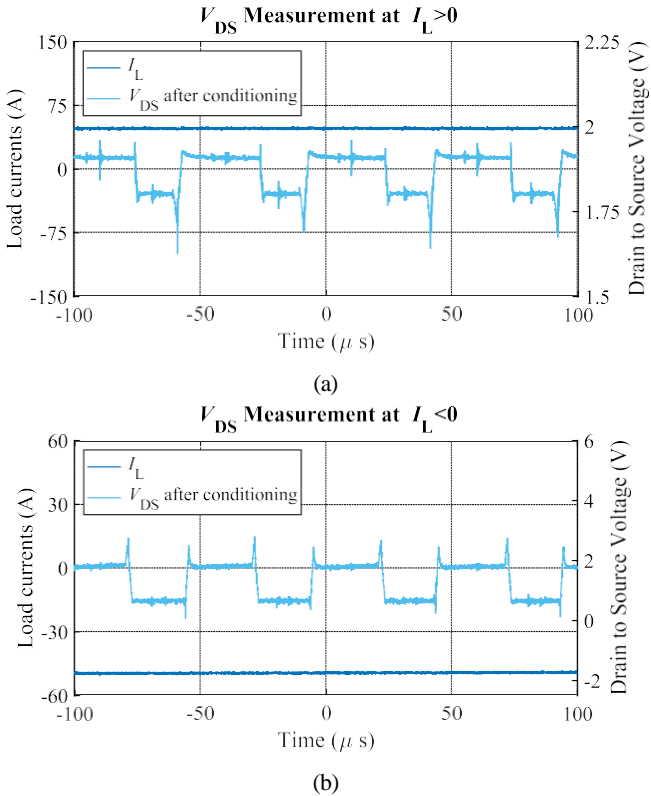

Fig. 9: Experimental waveforms of the drain to source measurement circuit. (a) $V_{DS}$ measurement at positive switch current for on-resistance monitoring. (b) $V_{DS}$ measurement at negative switch current for body diode voltage monitoring.

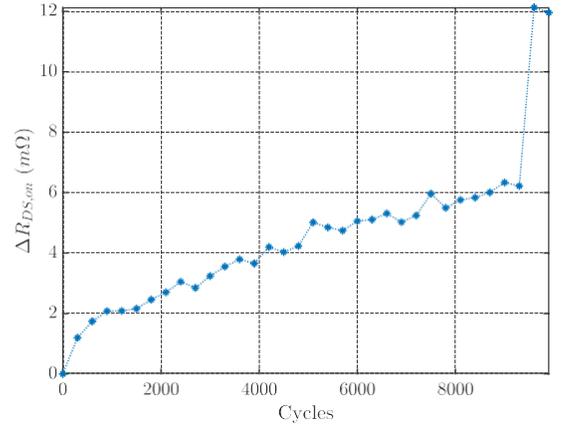

Fig. 10: On-resistance variation during aging.

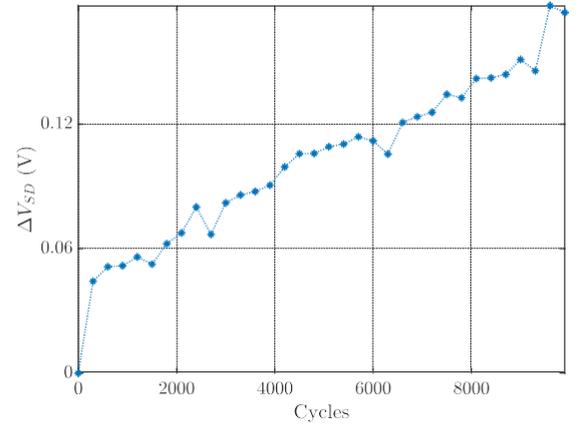

Fig. 11: Body diode voltage variation during aging.

When the switch operates at third-quadrant mode, a small part of the current flows through the $D_{CLP}$, $R_S$, $D_{HV1}$, and $D_{HV2}$. Therefore, the $V_{OP1}$ equals body diode voltage. Fig. 9 shows the experimental waveforms of the drain-source measurement circuit for on-resistance and body diode voltage monitoring. Figs. 10 and 11 illustrate the aging effect on the on-resistance and body diode voltage for the device under test, respectively.

B. Threshold Voltage Measurement

Fig. 12 shows the threshold voltage measurement circuit and its operational modes. The circuit is developed based on previous designs for IGBT [78] and GaN [79]. This circuit is also integrated into the DESAT protection circuit, reducing the number of components in the gate drive board. The threshold voltage can be measured shortly after the on-state. The circuit has two operating modes and is designed for in-situ monitoring. This circuit uses the internal chip sensor to measure the ambient temperature. Then, it measures the threshold voltage to avoid high junction temperature effect compensation. Since oxide aging is a slow phenomenon, this infrequent measurement is acceptable. First, the switch's gate-source capacitance is charged by DESAT current source. When the gate-source voltage reaches to the threshold value, the channel starts to conduct, and current flows through $R_S$, $D_{HV1}$, $D_{HV2}$, and the switch channel. The current source in the DESAT protection is typically adjustable. In this setup, the threshold voltage is monitored at $I_{DESAT} = 2mA$. Therefore, the injected current is low, and generated heat due to the conduction losses is negligible. There is a two-milli-second blanking time before measurement to ensure that the system reaches to the steady-state operation. The steady-state value is captured and transferred through the SPI interface to the microcontroller. In the steady-state, the leakage current is negligible, and all current flows through the channel. The device threshold voltage is changed by increasing the device temperature to check the circuit's efficacy. The threshold voltage is measured by both circuit and Keysight 1506A curve tracer. The results comparison confirms the error of less than 0.1 $V$. Fig. 13 shows the experimental waveforms of the threshold voltage monitoring circuit at different junction temperatures. The aging impact on the threshold voltage at the channel current of 2 $mV$ is depicted in Fig. 14.

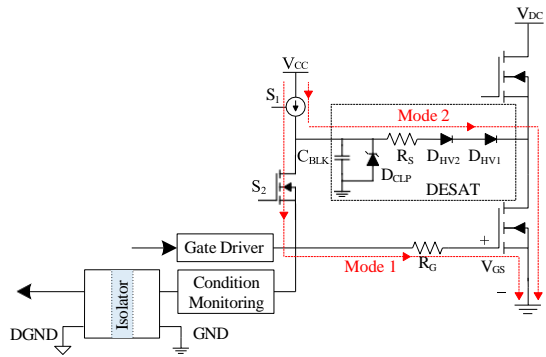

Fig. 12: Threshold voltage measurement circuit and its operational modes (Mode 1: Gate charging. Mode 2: Channel conduction).

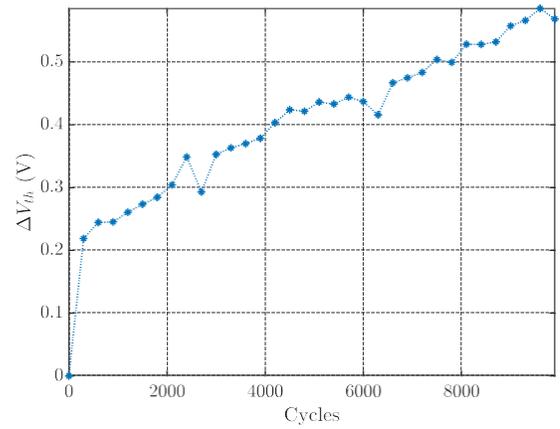

Fig. 14: Threshold voltage variation during aging.

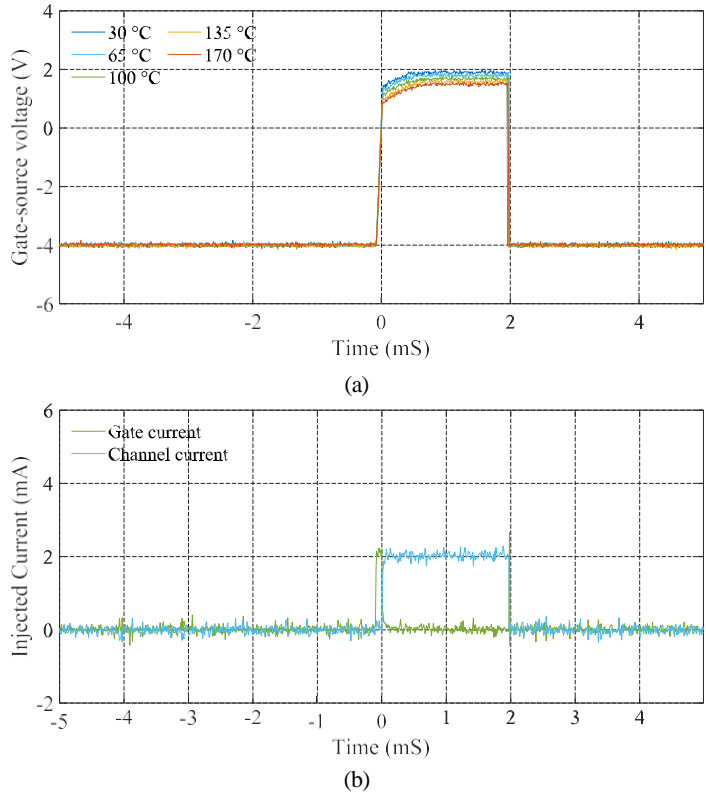

Fig. 13: Experimental waveforms of the threshold voltage monitoring circuit at different junction temperatures. (a) Gate-source voltage. (b) Gate current at voltage ramp-up mode and channel current at conduction mode.

## VI. JUNCTION TEMPERATURE MEASUREMENT

The accurate junction temperature monitoring is imperative for thermal cycling, condition monitoring, lifetime extension, loss management, and temperature protection. Therefore, sufficient accuracy for junction temperature measurement can improve state-of-health monitoring, lifetime models, and advanced converter applications. Junction temperature can be monitored directly by optical sensors [80] and integrated on-chip sensors [81]. However, these techniques are considered invasive and expensive. Another method is to use the TSEPs. Using electrical parameters for temperature measurement is different from aging monitoring. This is because aging is a slow phenomenon while the junction temperature can rapidly change. Therefore, selected electrical parameters for temperature measurement should be feasible for online monitoring. This section evaluates the TSEPs for online junction temperature measurement of SiC MOSFETs. Next, the practical implementation of the selected TSEP is discussed. The out-of-order equivalent time sampling technique is used to decrease the processing burden and avoid interference with core tasks in the code. Also, FIR filter is applied to the samples to increase the accuracy of measurements.

Table I summarizes the TSEPs for junction temperature measurement of the SiC MOSFETs. The on-resistance has relatively high sensitivity and can be easily implemented for online monitoring. However, the on-resistance relation with junction temperature is not perfectly linear in SiC MOSFETs. It is partially because the channel mobility increases over temperature and the channel resistance shows a negative temperature coefficient [98]. On the other hand, drift region mobility decreases as the temperature increases. Therefore, drift region resistance increases at elevated temperatures [99]. The opposite effect of temperature on channel and drift region resistances makes on-resistance a nonlinear temperature-dependent parameter. The new generation of SiC devices is designed to have a good trade-off between the channel and drift region resistance to achieve linear temperature dependency. The threshold voltage is a linear temperature-dependent parameter. However, a fast-sampling circuit is required for online threshold voltage measurements. In [83], the parasitic impedance between the kelvin source and power source pins is used to measure the threshold voltage, which is suitable only for 4-pin devices. Similarly, body diode voltage has a linear dependency on temperature. The package degradation effect on the body diode is significant at high currents. This makes the body diode voltage unsuitable to use at higher currents. However, the package degradation at lower currents does not affect the body diode voltage. The body diode voltage at lower current is highly sensitive and linear yet hard to implement online. In [71], [87]–[89], the internal gate resistance-based temperature measurements are proposed. This parameter is highly sensitive to noise and has relatively low linearity. The switching parameters like turn-on delay time [90]–[92], turn-on di/dt [83], [93]–[95], turn-off delay time [70], [96], [97], and turn-off di/dt [95], are presented in the literature for junction temperature measurement of SiC MOSFETs. A large gate resistance insertion during measurement is necessary to achieve sufficient accuracy for these parameters. Also, the PCB layout design can affect the switching parameters, and pre-start calibration is a cumbersome process for these parameters.

TABLE I: Summary of TSEPs for SiC MOSFETs junction temperature measurement.

| TSEP | Linearity | Sensitivity | Real-Time Implementation | Properties |
|---|---|---|---|---|
| On-Resistance, ($R_{ds,ON}$) [82]–[84] | Medium | 2.4 m$\Omega$/°C for vendor A<br>1.6 m$\Omega$/°C for vendor B | Easy | - Needs synchronized measurement for drain-source voltage and drain current. |
| Threshold Voltage, ($V_{th}$) [77], [83], [85] | Good | -6.4 mV/°C for vendor A<br>-3.1 mV/°C for vendor B | Hard | - Suitable for in-situ measurement. |
| Body Diode at High Current, ($V_{SD}$) [68], [85] | Good | -4.8 mV/°C | Easy | - Highly aging sensitive at high current. Affected by many electrical parameters. |
| Body Diode at Low Current, ($V_{SD}$) [68], [85] | Good | -2.65 mV/°C | Hard | - Needs high resolution low current measurement circuit.<br>- Monitoring circuits are In-situ. |
| Shoot-Through Current, ($I_{Sht}$) [86] | Medium | 0.5% /°C | Hard | - Depends on gate driver design |
| Gate Turn-On Peak Current, ($I_{G,peak}$) [71], [87]–[89] | Medium | 2.72 mA/°C | Hard | - Good for slow switching application. |
| Internal Gate Resistance, ($R_{G,int}$) [71], [87]–[89] | Medium | 3.6 m$\Omega$/°C | Medium | - Highly susceptible to noise. |
| Turn-On Delay Time, ($T_{d,on}$) [90]–[92] | Good | $\approx$ 1 nS/°C at $R_g$ = 2000$\Omega$ | Hard | - A large gate resistance needs to achieve detectable resolution. |
| Turn-On $di/dt$, [83], [93]–[95] | Low | 0.04 A/$\mu$S per °C at $R_g$ = 10$\Omega$ | Hard | - Needs very high detection resolution and accuracy. |
| Turn-Off Delay Time, ($T_{d,off}$) [70], [96], [97] | Good | 0.012 nS/°C at $R_g$ = 0$\Omega$<br>0.69 nS/°C at $R_g$ = 300$\Omega$ | Hard | - A large gate resistance needs to achieve detectable resolution. |
| Turn-Off $di/dt$, [95] | Low | 0.35 A/$\mu$S per °C at $R_g$ = 10$\Omega$ | Hard | - Needs very high detection resolution and accuracy. |

Considering linearity, sensitivity, and ease of real-time implementation, the on-resistance is a good candidate for online junction temperature monitoring. In this case, it's also beneficial as the online on-resistance measurement circuit is already available for aging monitoring. However, the on-resistance is package and gate-oxide degradation dependent. Therefore, the on-resistance and threshold voltage are periodically measured at start-up, and aging effect is compensated in the lookup table of on-resistance versus junction temperature and load current. Fig. 15 shows the on-resistance shift of a SiC power module as a function of junction temperature and drain current. As evident from the observations, the temperature dependence pattern exhibits a consistent shift across temperature and load conditions over aging. This shift can be quantified by recalibrating the temperature lookup table when initiating the test and when the devices are at ambient temperature.

Several on-resistance measurement circuits are proposed in the literature. However, they didn't discuss executing all the necessary steps for online on-resistance measurement. Implementing on-resistance measurement with microcontroller is an overlooked challenging task in the condition monitoring due to the data acquisition limitation. This setup monitors twelve on-state voltages and three load currents online. Therefore, a code-efficient data acquisition technique is crucial to avoid condition monitoring interference with core code. Also, the data acquisition technique should be implementable on a usual microcontroller with a sufficient sampling rate. The core code consists of control loop execution, PWM update, and ADC reading, which runs in the main interrupt service routine (ISR) with the highest priority. The other tasks like SPI interface and condition monitoring are run in the secondary and tertiary ISRs. These tasks can be executed in several periods. Since the on-state voltage and load currents are not perfectly periodic, it is essential to minimize the number of periods for each condition monitoring. Sequential equivalent time sampling is the straightforward method to measure N data points at N cycles. The out-of-order equivalent time sampling is used to decrease the number of required cycles for N data points measurement. Fig. 16 shows the out-of-order equivalent time sampling of load current and on-state voltage for on-resistance measurement. This method captures several data points at different cycles without order. To this end, the electrical angle is used as the reference, and its corresponding values for specific on-state voltage phases are determined. To increase the accuracy, the electrical angle corresponding to peak values of load current and on-state voltage is used to measure the on-resistance. More specifically, 300 data points around the peak values are selected. If the electrical angle in execution ISR is equal to any 300 selected data points, the current and on-state voltage values are immediately captured and logged to the array. The recursive binary search technique is used to optimize the searching for selected data points. Fig. 17 (a), (b) shows the experimental results of the 300 on-state voltage and load current measurements around the peak values. The measured on-resistance is susceptible to noise, and data filtering is needed for accurate condition monitoring. A FIR filter is useful to filter the on-resistance raw data. Fig. 17 (c) shows the experimental results of the filtered on-resistance measurements. The filtered results are compared with reference values from the Keysight 1506A curve tracer. It is shown that there is an error of less than 1.5 % for online on-resistance measurements.

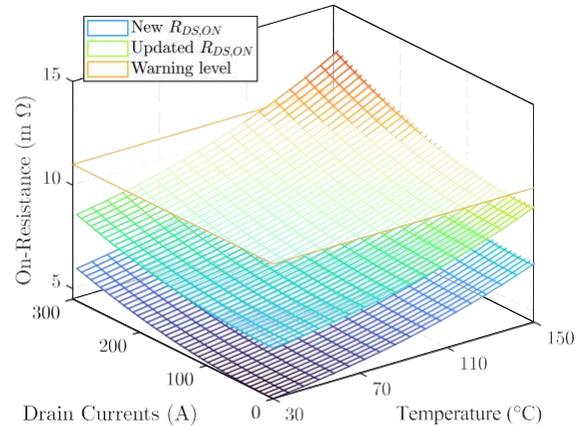

Fig. 15: On-resistance of a SiC power module as a function of junction temperature and drain current.

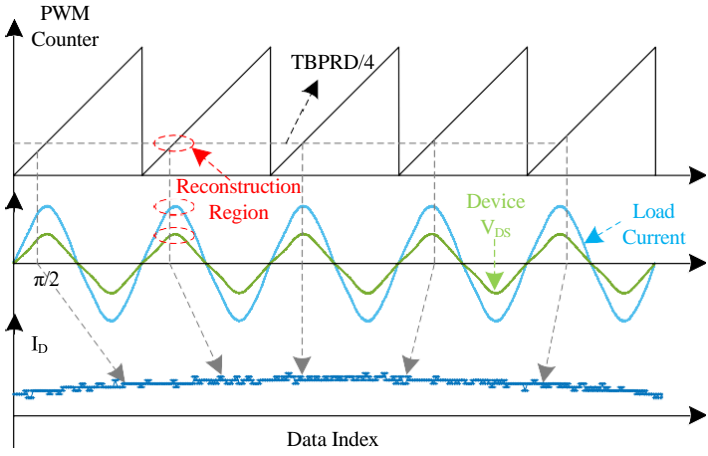

Fig. 16: Current and voltage sampling technique for on-resistance measurement.

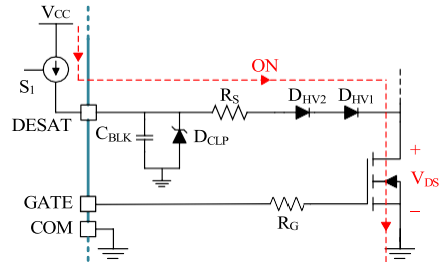

Fig. 18: Circuit diagram of DESAT for overcurrent and short circuit protection.

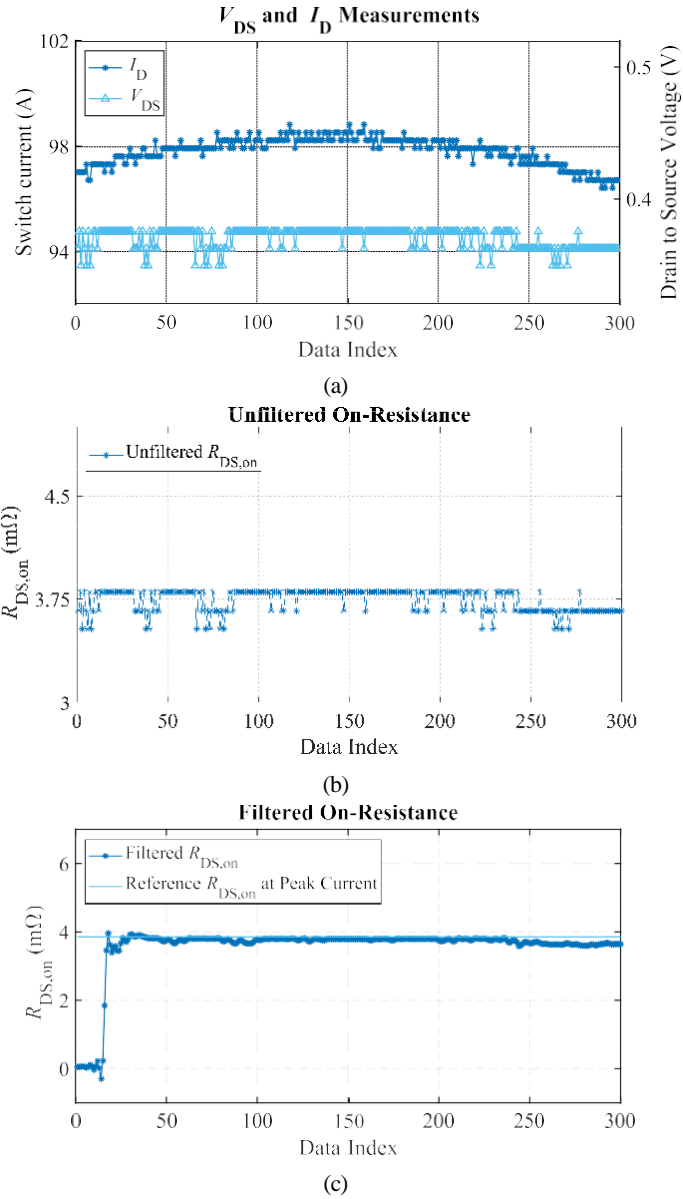

Fig. 17: Experimental results of the 300 on-resistance measurements around the peak current. (a) On-state voltage and load current results. (b) Unfiltered on-resistance results. (c) Filtered on-resistance results.

## VII. SETUP DESIGN CONSIDERATIONS

### A. Transient Overvoltage Protection

The short-circuit events at the end of the device lifetime can trigger the protection circuits. The protection circuit isolates the rest of the devices by turning them off. However, the inductor currents flow through body diodes and charge the DC link capacitors. This can generate a transient high voltage in DC bus voltage and damage devices and DC link capacitors. The metal oxide varistor (MOV) is connected in parallel to the DC link capacitors. The MOV is suitable protection for transient voltages due to its fast response and high power capacity. The selected MOV should have 1) a higher steady-state voltage rating than the DC bus voltage, 2) a lower voltage clamping than the maximum voltage of power modules, and 3) sufficient transient energy to absorb load inductors and DC link capacitors.

### B. Desaturation Detection Circuit

DESAT protection is the primary response to overcurrent and short-circuit faults. Fig. 18 shows the circuit diagram of DESAT.

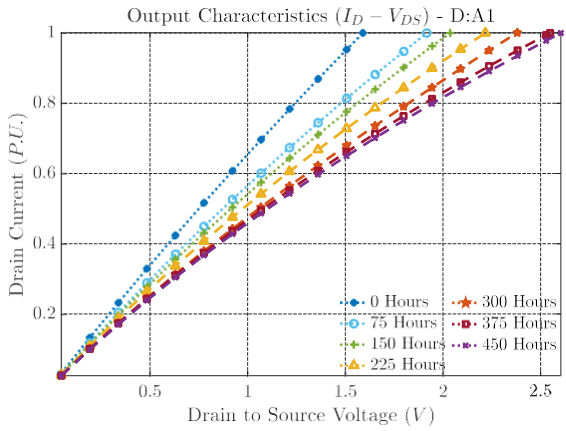

Fig. 19: Experimental results of aging effect on drain to source voltage at $V_{gs} = 15V$.

During the switch turn-on, a constant current flows through $R_S$, $D_{HV\,1}$, $D_{HV\,2}$, and the switch channel. The voltage at DESAT pin can be calculated as:

$$V_{DESAT} = I_{DESAT} \times R_S + V_{DHV,1} + V_{DHV,2} + V_{DS} \quad (7)$$

In (7), $V_{DS}$ is a function of drain current. If the drain current exceeds the current value corresponding to saturation threshold voltage for longer than a fixed blanking time, the DESAT protection feature is triggered.

The output characteristic of the device is highly sensitive to gate-oxide degradation. Fig. 19 shows the experimental results of the oxide aging effect on the drain to source voltage at $V_{gs} = 15V$. As can be seen in this figure, $V_{DS}$ increases from 1.58 V to 2.6 V for nominal current. This can lead to false triggers in DESAT circuit. To avoid this, threshold voltage shift can be used as a criterion for gate oxide aging and adjust the threshold voltage for DESAT accordingly.

It should be noted that the DESAT circuit is susceptible to noise in high-voltage applications. To reduce the noise effect, the composed RC filter of $C_{BLK}$, $R_S$, and $D_{CLP}$ are placed as close as possible to the DESAT pin in the PCB layout. Also, DESAT path is placed in an inner layer between two reference planes.

### C. Common Mode Noise

Signal integrity is one of the main challenges in the design of the gate driver and main controller boards. Due to the high dv/dt of switching transients for SiC MOSFETs, electromagnetic interference (EMI) and unwanted noises can easily affect the signals and cause malfunction. In this test setup, each gate driver has its own ground for measurement and isolated power supply for device driving. However, the gate driver ICs and isolated DC power supplies have 2 $pF$ and 4 $pF$ coupling capacitors, respectively. These coupling capacitors form common mode noise paths in the setup. Fig 20 (a) illustrates the common-mode noise paths for one leg of AC power cycling test setup. A considerable common mode current flows the coupling capacitors during the switching transients. One of the precautions to reduce the common mode noise is inserting the common mode chokes in the noise paths. Fig 20 (b) illustrates the common-mode (CM) choke insertion to suppress the noise. Ferrite common mode chokes insert high-frequency resistors and minimize the common mode noise effect in the circuit. Fig. 21 shows the experimental waveforms of the common-mode choke effect on CM current of the power supply and the SPI clock signal. It is shown that the current peaks at the common mode current of the DC power supply are eliminated. Also, the common mode noise interference with other signals like the SPI clock is minimized.

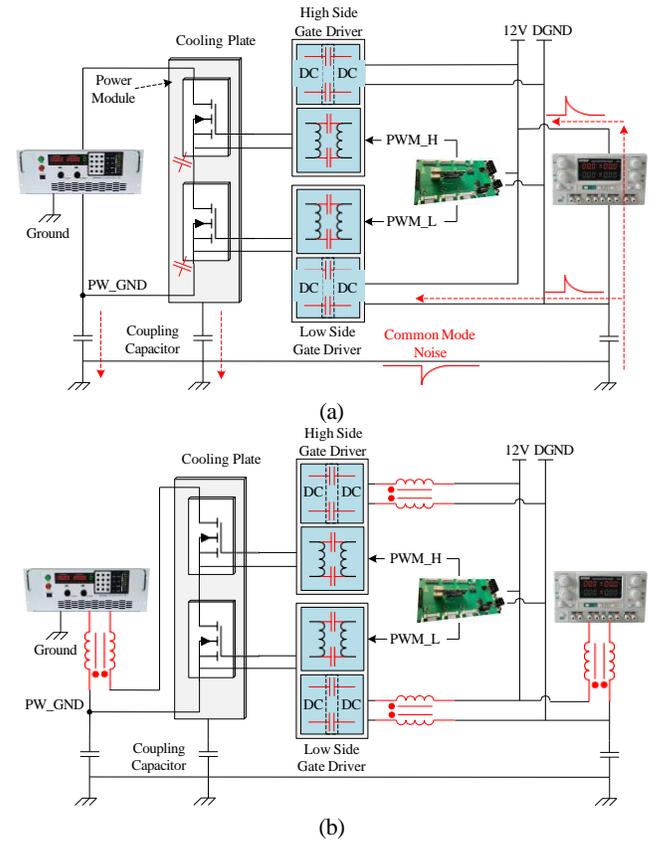

Fig. 20: Illustration of the common-mode noise in AC power cycling test setup. (a) Common-mode noise paths for one leg of the AC power cycling test setup. (b) Common-mode choke insertion to suppress the noise.

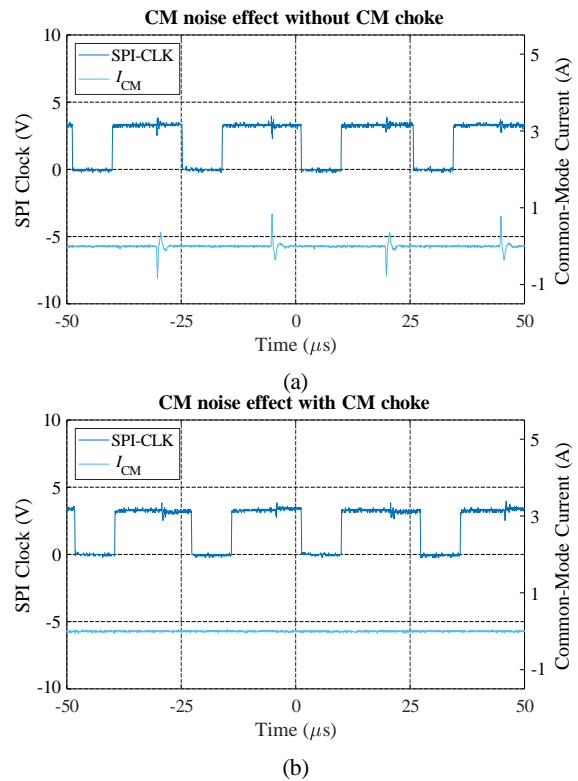

Fig. 21: Experimental waveforms of the common-mode choke effect on CM current of the power supply and the SPI clock signal. (a) Circuit without CM choke. (b) Circuit with CM choke.

From the PCB point of view, tracks for high-speed logic like PWMs and SPI signals are routed inside inner reference voltage planes to shield them against spurious signals and emissions. Also, the termination CRC filter is added for these signals between the gate driver and main board to increase noise immunity.

## VIII. Conclusion

This paper presents an insightful summary of the degradation mechanisms and accelerated lifetime tests for SiC MOSFETs which became very popular in the EV and renewable energy market. A 120 KVA AC power cycling test setup is developed with online condition monitoring circuits to cover all dominant failure mechanisms to report impending failures to users in advance. AC power cycling test setup configuration and control method are discussed in detail. Also, a comparison among the available precursors in the literature is presented to identify the suitable precursors for condition monitoring. The on-resistance, body diode voltage, and threshold voltage are selected as the proper precursors, and corresponding condition monitoring circuits are developed. The condition monitoring circuits are combined with DESAT protection circuits in the EV gate driver to minimize the number of components and make them easy to implement. The practical challenges in the implementation of monitoring circuits are identified and addressed. The TSEPs for SiC MOSFETs are compared, and the best option is highlighted for junction temperature measurement. An out-of-order equivalent time sampling algorithm is developed for data logging to avoid a high data acquisition rate and interference with core code tasks. This data sampling technique efficiently decreased the number of required cycles. It is shown that the filtered online on-resistance measurements have an error of less than 1.5 % compared to the Keysight 1506A curve tracer results. Finally, the design considerations for practical issues are analyzed. Specifically, the common-mode noise and aging effect on DESAT protection are discussed in depth.

## Acknowledgement

The authors would like to thank Texas Instrument Automotive Group and Semiconductor Research Corporation (SRC)/Texas Analog Center of Excellence (TxACE) for supporting this research under Task ID 2810.054.

## References

[1] F. Meinguet, P. Sandulescu, X. Kestelyn, and E. Semail, "A Method for Fault Detection and Isolation Based on the Processing of Multiple Diagnostic Indices: Application to Inverter Faults in AC Drives," *IEEE Transactions on Vehicular Technology*, vol. 62, no. 3, pp. 995–1009, 2013.

[2] K. Li, S. Cheng, T. Yu, X. Wu, C. Xiang, and A. Bilal, "An On-Line Multiple Open-Circuit Fault Diagnostic Technique for Railway Vehicle Air-Conditioning Inverters," *IEEE Transactions on Vehicular Technology*, vol. 69, no. 7, pp. 7026–7039, 2020.

[3] J. Zhang, H. Yao, and G. Rizzoni, "Fault Diagnosis for Electric Drive Systems of Electrified Vehicles Based on Structural Analysis," *IEEE Transactions on Vehicular Technology*, vol. 66, no. 2, pp. 1027–1039, 2017.

[4] S. Yang, A. Bryant, P. Mawby, D. Xiang, L. Ran, and P. Tavner, "An Industry-Based Survey of Reliability in Power Electronic Converters," *IEEE Transactions on Industry Applications*, vol. 47, no. 3, pp. 1441–1451, 2011.

[5] B. Ozpineci and L. M. Tolbert, *Comparison of Wide-Bandgap Semiconductors for Power Electronics Applications*. United States. Department of Energy, 2004.

[6] K. Boomer, J.-M. Lauenstein, and A. Hammoud, "Body of Knowledge for Silicon Carbide Power Electronics," 2016. [Online]. Available: https://ntrs.nasa.gov/archive/nasa/casi.ntrs.nasa.gov/20170003922.pdf

[7] R. Mousa, "Characterization, Modelling and Integration of Silicon Carbide Power JFET in High Temperature High Voltage Converters," *PhD Thesis*, pp. 1–169, 2009. [Online]. Available: http://www.theses.fr/2009ISAL0043

[8] X. Ding, M. Du, C. Duan, H. Guo, R. Xiong, J. Xu, J. Cheng, and P. Chi kwong Luk, "Analytical and Experimental Evaluation of SiC-Inverter Nonlinearities for Traction Drives Used in Electric Vehicles," *IEEE Transactions on Vehicular Technology*, vol. 67, no. 1, pp. 146–159, 2018.

[9] N. Kaminski, "State of The Art and The Future of Wide Band-Gap Devices," in *2009 13th European Conference on Power Electronics and Applications*, pp. 1–9, 2009.

[10] "C3M0075120K data sheet," Cree, Inc, Durham, North Carolina, USA.

[11] "APT28M120B2 data sheet," Microsemi, Inc, Culver City, California, USA.

[12] K. Suganuma, *Wide Bandgap Power Semiconductor Packaging: Materials, Components, and Reliability*, ser. Woodhead Publishing Series in Electronic and Optical Materials. Woodhead Publishing, 2018.

[13] K. O. Dohnke, K. Guth, and N. Heuck, "History and Recent Developments of Packaging Technology for SiC Power Devices," in *Silicon Carbide and Related Materials 2015*, ser. Materials Science Forum, vol. 858, pp. 1043–1048. Trans Tech Publications Ltd, 6 2016.

[14] T. Kimoto, "Bulk and Epitaxial Growth of Silicon Carbide," *Progress in Crystal Growth and Characterization of Materials*, vol. 62, no. 2, pp. 329–351, 2016.

[15] O. A. Salvado, "Contribution to The Study of The SiC MOSFETs Gate Oxide," *PhD Thesis*, 2018.

[16] J. Wang and X. Jiang, "Review and Analysis of SiC MOSFETs' Ruggedness and Reliability," *IET Power Electronics*, vol. 13, no. 3, pp. 445–455, 2020.

[17] T. Rahimi and et al., "Fuzzy Lifetime Analysis of a Fault-Tolerant Two-Phase Interleaved Converter," in *2021 IEEE 13th International Symposium on Diagnostics for Electrical Machines, Power Electronics and Drives (SDEMPED)*, vol. 1, pp. 407–412, 2021.

[18] T. Rahimi and et al., "Fault-Tolerant Performance Enhancement of DC-DC Converters with High-Speed Fault Clearing-unit based Redundant Power Switch Configurations," in *2021 IEEE Electrical Power and Energy Conference (EPEC)*, pp. 492–497, 2021.

[19] M. Andresen and M. Liserre, "Impact of Active Thermal Management on Power Electronics Design," *Microelectronics Reliability*, vol. 54, no. 9, pp. 1935–1939, 2014.

[20] T. Rahimi and et al., "Unbalanced Currents Effect on the Thermal Characteristic and Reliability of Parallel Connected Power Switches," *Case Studies in Thermal Engineering*, vol. 26, p. 101134, 2021.

[21] S. Xu, W. Huang, H. Wang, W. Zheng, J. Wang, Y. Chai, and M. Ma, "A Simultaneous Diagnosis Method for Power Switch and Current Sensor Faults in Grid-Connected Three-Level NPC Inverters," *IEEE Transactions on Power Electronics*, vol. 38, no. 1, pp. 1104–1118, 2023.

[22] B. T. Vankayalapati, S. Pu, F. Yang, M. Farhadi, V. Gurusamy, and B. Akin, "Investigation and On-Board Detection of Gate-Open Failure in SiC MOSFETs," *IEEE Transactions on Power Electronics*, vol. 37, no. 4, pp. 4658–4671, 2022.

[23] B. T. Vankayalapati, M. Farhadi, R. Sajadi, B. Akin, and H. Tan, "A Practical Switch Condition Monitoring Solution for SiC Traction Inverters," *IEEE Journal of Emerging and Selected Topics in Power Electronics*, 2022.

[24] W. Xu, S. Qu, and C. Zhang, "Fast Terminal Sliding Mode Current Control With Adaptive Extended State Disturbance Observer for PMSM System," *IEEE Journal of Emerging and Selected Topics in Power Electronics*, vol. 11, no. 1, pp. 418–431, 2023.

[25] C. Lu, R. Zhu, F. Yu, X. Jiang, Z. Liu, L. Dong, Q. Hua, and Z. Ou, "Gear rotational speed sensor based on fecosib/pb(zr,ti)o3 magnetoelectric composite," *Measurement*, vol. 168, p. 108409, 2021.

[26] S. Liu, Z. Song, Z. Dong, Y. Liu, and C. Liu, "Generic Carrier-Based PWM Solution for Series-End Winding PMSM Traction System With Adaptative Overmodulation Scheme," *IEEE Transactions on Transportation Electrification*, vol. 9, no. 1, pp. 712–726, 2023.

[27] S. Liu, Z. Song, Y. Liu, Y. Chen, and C. Liu, "Flux-Weakening Controller Design of Dual Three-Phase PMSM Drive System With Copper Loss Minimization," *IEEE Transactions on Power Electronics*, vol. 38, no. 2, pp. 2351–2363, 2023.

[28] M. Farhadi, M. Abapour, and B. Mohammadi-Ivatloo, "Reliability Analysis of Component-Level Redundant Topologies for Solid-State Fault Current Limiter," *International Journal of Electronics*, vol. 105, no. 4, pp. 541–558, 2018.

[29] M. Farhadi, M. T. Fard, M. Abapour, and M. T. Hagh, "DC-AC Converter-Fed Induction Motor Drive With Fault-Tolerant Capability Under Open- and

Short-Circuit Switch Failures," *IEEE Transactions on Power Electronics*, vol. 33, no. 2, pp. 1609–1621, 2018.

[30] M. Farhadi, M. Abapour, and M. Sabahi, "Failure Analysis and Reliability Evaluation of Modulation Techniques for Neutral Point Clamped Invertersâ A Usage Model Approach," *Engineering Failure Analysis*, vol. 71, pp. 90–104, 2017.

[31] B. T. Vankayalapati, F. Yang, S. Pu, M. Farhadi, and B. Akin, "A Highly Scalable, Modular Test Bench Architecture for Large-Scale DC Power Cycling of SiC MOSFETs: Towards Data Enabled Reliability," *IEEE Power Electronics Magazine*, vol. 8, no. 1, pp. 39–48, 2021.

[32] "Temperature, Bias, and Operating Life," *JEDEC Standard JESD22-A108D*, 2010.

[33] L. Yang and A. Castellazzi, "High Temperature Gate-Bias and Reverse-Bias Tests on SiC MOSFETs," *Microelectronics Reliability*, vol. 53, no. 9, pp. 1771–1773, 2013.

[34] X. Jiang, J. Wang, J. Chen, H. Yu, Z. Li, and Z. John Shen, "Investigation on Degradation of SiC MOSFET Under Accelerated Stress in a PFC Converter," *IEEE Journal of Emerging and Selected Topics in Power Electronics*, vol. 9, no. 4, pp. 4299–4310, 2021.

[35] V. Smet, F. Forest, J.-J. Huselstein, F. Richardeau, Z. Khatir, S. Lefebvre, and M. Berkani, "Ageing and Failure Modes of IGBT Modules in High-Temperature Power Cycling," *IEEE Transactions on Industrial Electronics*, vol. 58, no. 10, pp. 4931–4941, 2011.

[36] C. H. van der Broeck, L. A. Ruppert, R. D. Lorenz, and R. W. De Doncker, "Methodology for Active Thermal Cycle Reduction of Power Electronic Modules," *IEEE Transactions on Power Electronics*, vol. 34, no. 8, pp. 8213–8229, 2019.

[37] U.-M. Choi, S. Jorgensen, and F. Blaabjerg, "Advanced Accelerated Power Cycling Test for Reliability Investigation of Power Device Modules," *IEEE Transactions on Power Electronics*, vol. 31, no. 12, pp. 8371–8386, 2016.

[38] R. Ruffilli, "Fatigue Mechanisms in Al-Based Metallizations in Power MOSFETs," Theses, Université Paul Sabatier (Toulouse 3) ; École doctorale Sciences de la Matière, Dec. 2017. [Online]. Available: https://hal.archives-ouvertes.fr/tel-01666437

[39] T. Ishigaki, T. Murata, K. Kinoshita, T. Morikawa, T. Oda, R. Fujita, K. Konishi, Y. Mori, and A. Shima, "Analysis of Degradation Phenomena in Bipolar Degradation Screening Process for SiC-MOSFETs," in *2019 31st International Symposium on Power Semiconductor Devices and ICs (ISPSD)*, pp. 259–262, 2019.

[40] X. Jiang, J. Wang, J. Lu, J. Chen, X. Yang, Z. Li, C. Tu, and Z. J. Shen, "Failure Modes and Mechanism Analysis of SiC MOSFET Under Short-Circuit Conditions," *Microelectronics Reliability*, vol. 88-90, pp. 593–597, 2018.

[41] S.-H. Tran, Z. Khatir, R. Lallemand, A. Ibrahim, J.-P. Ousten, J. Ewanchuk, and S. V. Mollov, "Constant $\Delta T j$ Power Cycling Strategy in DC Mode for Top-Metal and Bond-Wire Contacts Degradation Investigations," *IEEE Transactions on Power Electronics*, vol. 34, no. 3, pp. 2171–2180, 2019.

[42] H. Luo, F. Iannuzzo, F. Blaabjerg, W. Li, and X. He, "Separation Test Method for Investigation of Current Density Effects on Bond Wires of SiC Power MOSFET Modules," in *IECON 2017 - 43rd Annual Conference of the IEEE Industrial Electronics Society*, pp. 1525–1530, 2017.

[43] C. S. Hau-Riege, "An Introduction to Cu Electromigration," *Microelectronics Reliability*, vol. 44, no. 2, pp. 195–205, 2004.

[44] H. Luo, F. Iannuzzo, N. Baker, F. Blaabjerg, W. Li, and X. He, "Study of Current Density Influence on Bond Wire Degradation Rate in SiC MOSFET Modules," *IEEE Journal of Emerging and Selected Topics in Power Electronics*, vol. 8, no. 2, pp. 1622–1632, 2020.

[45] B. Hu, S. Konaklieva, N. Kourra, M. A. Williams, L. Ran, and W. Lai, "Long-Term Reliability Evaluation of Power Modules With Low Amplitude Thermomechanical Stresses and Initial Defects," *IEEE Journal of Emerging and Selected Topics in Power Electronics*, vol. 9, no. 1, pp. 602–615, 2021.

[46] S. Singh, J. Hao, D. Hoffman, T. Dixon, A. Zedolik, J. Fazio, and T. E. Kopley, "Effects of Die-Attach Voids on the Thermal Impedance of Power Electronic Packages," *IEEE Transactions on Components, Packaging and Manufacturing Technology*, vol. 7, no. 10, pp. 1608–1616, 2017.

[47] M. Ciappa, "Selected Failure Mechanisms of Modern Power Modules," *Microelectronics Reliability*, vol. 42, no. 4, pp. 653–667, 2002.

[48] A. Sokolov, C. Liu, and F. Mohn, "Reliability Assessment of SiC Power Module Stack Based on Thermo-Structural Analysis," in *2018 19th International Conference on Thermal, Mechanical and Multi-Physics Simulation and Experiments in Microelectronics and Microsystems (EuroSimE)*, pp. 1–4, 2018.

[49] P. Friedrichs, "High-Performance SiC MOSFET Technology for Power Electronics Design," Nov. 2019. [Online]. Available: http://www.infineon.com

[50] V. Afanasev, M. Bassler, G. Pensl, and M. Schulz, "Intrinsic SiC/SiO2 Interface States," *Physica Status Solidi (A) Applied Research*, vol. 162, no. 1, pp. 321–337, 1997.

[51] C. J. Cochrane, P. M. Lenahan, and A. J. Lelis, "An Electrically Detected Magnetic Resonance Study of Performance Limiting Defects in SiC Metal Oxide Semiconductor Field Effect Transistors," *Journal of Applied Physics*, vol. 109, no. 1, p. 014506, 2011.

[52] R. Singh, "Reliability and Performance Limitations in SiC Power Devices," *Microelectronics Reliability*, vol. 46, no. 5, pp. 713–730, 2006.

[53] L. C. Yu, K. P. Cheung, J. S. Suehle, J. P. Campbell, K. Sheng, A. J. Lelis, and S. H. Ryu, "Channel Hot-Carrier Effect of 4H-SiC MOSFET," in *Silicon Carbide and Related Materials 2008*, ser. Materials Science Forum, vol. 615, pp. 813–816, 7 2009.

[54] T. Funaki, "A Study on Performance Degradation of SiC MOSFET for Burn-In Test of Body Diode," in *2013 4th IEEE International Symposium on Power Electronics for Distributed Generation Systems (PEDG)*, pp. 1–5, 2013.

[55] J. Q. Liu, M. Skowronski, C. Hallin, R. SOderholm, and H. Lendenmann, "Structure of Recombination-Induced Stacking Faults in High-Voltage SiC p-n Junctions," *Applied Physics Letters*, vol. 80, no. 5, pp. 749–751, 2002.

[56] M. Skowronski and S. Ha, "Degradation of Hexagonal Silicon-Carbide-Based Bipolar Devices," *Journal of Applied Physics*, vol. 99, no. 1, 2006.

[57] K. Konishi, R. Fujita, A. Shima, and Y. Shimamoto, "Modeling of Stacking Fault Expansion Velocity of Body Diode in 4H-SiC MOSFET," in *2016 European Conference on Silicon Carbide Related Materials (ECSCRM)*, pp. 1–1, 2016.

[58] U. Karki and F. Z. Peng, "Precursors of Gate-Oxide Degradation in Silicon Carbide MOSFETs," in *2018 IEEE Energy Conversion Congress and Exposition (ECCE)*, pp. 857–861, 2018.

[59] A. Fayyaz and A. Castellazzi, "High Temperature Pulsed-Gate Robustness Testing of SiC Power MOSFETs," *Microelectronics Reliability*, vol. 55, no. 9, pp. 1724–1728, 2015.

[60] R. Bayerer, T. Herrmann, T. Licht, J. Lutz, and M. Feller, "Model for Power Cycling lifetime of IGBT Modules - Various Factors Influencing Lifetime," in *5th International Conference on Integrated Power Electronics Systems*, pp. 1–6, 2008.

[61] "ECPE Guideline AQG 324 - Qualification of Power Modules for Use in Power Electronic Converter units in Motor Vehicles," *Release 2.1/2019*, 2019.

[62] O. Avino-Salvado, C. Cheng, C. Buttay, H. Morel, D. Labrousse, S. Lefebvre, and M. Ali, "SiC MOSFETs Robustness for Diode-Less Applications," *EPE Journal*, vol. 28, no. 3, pp. 128–135, 2018.

[63] A. Bolotnikov, J. Glaser, J. Nasadoski, P. Losee, S. Klopman, A. Permuy, and L. Stevanovic, "Utilization of SiC MOSFET Body Diode in Hard Switching Applications," in *Silicon Carbide and Related Materials 2013*, vol. 778, pp. 947–950, 5 2014.

[64] T. Aichinger, G. Rescher, and G. Pobegen, "Threshold Voltage Peculiarities and Bias Temperature Instabilities of SiC MOSFETs," *Microelectronics Reliability*, vol. 80, pp. 68–78, 2018.

[65] K. Puschkarsky, H. Reisinger, T. Aichinger, W. Gustin, and T. Grasser, "Understanding BTI in SiC MOSFETs and Its Impact on Circuit Operation," *IEEE Transactions on Device and Materials Reliability*, vol. 18, no. 2, pp. 144–153, 2018.

[66] R. Bonyadi, O. Alatise, S. Jahdi, J. Hu, L. Evans, and P. A. Mawby, "Investigating the Reliability of SiC MOSFET Body Diodes Using Fourier Series Modelling," in *2014 IEEE Energy Conversion Congress and Exposition (ECCE)*, pp. 443–448, 2014.

[67] A. Fayyaz, G. Romano, and A. Castellazzi, "Body Diode Reliability Investigation of SiC Power MOSFETs," *Microelectronics Reliability*, vol. 64, pp. 530–534, 2016, proceedings of the 27th European Symposium on Reliability of Electron Devices, Failure Physics and Analysis.

[68] E. Ugur, C. Xu, F. Yang, S. Pu, and B. Akin, "A New Complete Condition Monitoring Method for SiC Power MOSFETs," *IEEE Transactions on Industrial Electronics*, vol. 68, no. 2, pp. 1654–1664, 2021.

[69] E. Ugur, F. Yang, S. Pu, S. Zhao, and B. Akin, "Degradation Assessment and Precursor Identification for SiC MOSFETs Under High Temp Cycling," *IEEE Transactions on Industry Applications*, vol. 55, no. 3, pp. 2858–2867, 2019.

[70] J. Dyer, Z. Zhang, F. Wang, D. Costinett, L. M. Tolbert, and B. J. Blalock, "Online Condition Monitoring of SiC Devices Using Intelligent Gate Drive for Converter Performance Improvement," in *2016 IEEE 4th Workshop on Wide Bandgap Power Devices and Applications (WiPDA)*, pp. 182–187, 2016.

[71] R. Wang, J. Sabate, K. Mainali, T. Sadilek, P. Losee, and Y. Singh, "SiC Device Junction Temperature Online Monitoring," in *2018 IEEE Energy Conversion Congress and Exposition (ECCE)*, pp. 387–392, 2018.

[72] U. Karki and F. Z. Peng, "Effect of Gate-Oxide Degradation on Electrical Parameters of Power MOSFETs," *IEEE Transactions on Power Electronics*, vol. 33, no. 12, pp. 10 764–10 773, 2018.

[73] F. Erturk, E. Ugur, J. Olson, and B. Akin, "Real-Time Aging Detection of SiC MOSFETs," *IEEE Transactions on Industry Applications*, vol. 55, no. 1, pp. 600–609, 2019.

[74] H. Luo, F. Iannuzzo, and M. Turnaturi, "Role of Threshold Voltage Shift in Highly Accelerated Power Cycling Tests for SiC MOSFET Modules," *IEEE Journal of Emerging and Selected Topics in Power Electronics*, vol. 8, no. 2, pp. 1657–1667, 2020.

[75] L. Ceccarelli, A. Bahman, and F. Iannuzzo, "Impact of Device Aging In The Compact Electro-Thermal Modeling of SiC Power MOSFETs," *Microelectronics Reliability*, vol. 100-101, p. 113336, 2019.

[76] S. Pu, F. Yang, B. T. Vankayalapati, E. Ugur, C. Xu, and B. Akin, "A Practical On-Board SiC MOSFET Condition Monitoring Technique for Aging Detection," *IEEE Transactions on Industry Applications*, vol. 56, no. 3, pp. 2828–2839, 2020.

[77] M. Farhadi, F. Yang, S. Pu, B. T. Vankayalapati, and B. Akin, "Temperature-Independent Gate-Oxide Degradation Monitoring of SiC MOSFETs Based on Junction Capacitances," *IEEE Transactions on Power Electronics*, vol. 36, no. 7, pp. 8308–8324, 2021.

[78] S. H. Ali, X. Li, A. S. Kamath, and B. Akin, "A Simple Plug-In Circuit for IGBT Gate Drivers to Monitor Device Aging: Toward Smart Gate Drivers," *IEEE Power Electronics Magazine*, vol. 5, no. 3, pp. 45–55, 2018.

[79] F. Yang, C. Xu, and B. Akin, "Characterization of Threshold Voltage Instability Under Off-State Drain Stress and Its Impact on p-GaN HEMT Performance," *IEEE Journal of Emerging and Selected Topics in Power Electronics*, vol. 9, no. 4, pp. 4026–4035, 2021.

[80] C. Li, H. Luo, C. Li, W. Li, H. Yang, and X. He, "Online Junction Temperature Extraction of SiC Power MOSFETs With Temperature Sensitive Optic Parameter (TSOP) Approach," *IEEE Transactions on Power Electronics*, vol. 34, no. 10, pp. 10 143–10 152, 2019.

[81] E. R. Motto and J. F. Donlon, "IGBT Module with User Accessible On-Chip Current and Temperature Sensors," in *2012 Twenty-Seventh Annual IEEE Applied Power Electronics Conference and Exposition (APEC)*, pp. 176–181, 2012.

[82] Y. Zhang and Y. C. Liang, "A Simple Approach On Junction Temperature Estimation for SiC MOSFET Dynamic Operation Within Safe Operating Area," in *2015 IEEE Energy Conversion Congress and Exposition (ECCE)*, pp. 5704–5707, 2015.

[83] A. Griffo, J. Wang, K. Colombage, and T. Kamel, "Real-Time Measurement of Temperature Sensitive Electrical Parameters in SiC Power MOSFETs," *IEEE Transactions on Industrial Electronics*, vol. 65, no. 3, pp. 2663–2671, 2018.

[84] H. Niu and R. D. Lorenz, "Sensing Power MOSFET Junction Temperature Using Circuit Output Current Ringing Decay," *IEEE Transactions on Industry Applications*, vol. 51, no. 2, pp. 1763–1773, 2015.

[85] F. Yang, E. Ugur, and B. Akin, "Evaluation of Aging's Effect on Temperature-Sensitive Electrical Parameters in SiC MOSFETs," *IEEE Transactions on Power Electronics*, vol. 35, no. 6, pp. 6315–6331, 2020.

[86] S. Jahdi, O. Alatise, J. A. Ortiz Gonzalez, R. Bonyadi, L. Ran, and P. Mawby, "Temperature and Switching Rate Dependence of Crosstalk in Si-IGBT and SiC Power Modules," *IEEE Transactions on Industrial Electronics*, vol. 63, no. 2, pp. 849–863, 2016.

[87] H. Niu and R. D. Lorenz, "Real-Time Junction Temperature Sensing for Silicon Carbide MOSFET With Different Gate Drive Topologies and Different Operating Conditions," *IEEE Transactions on Power Electronics*, vol. 33, no. 4, pp. 3424–3440, 2018.

[88] H. Niu and R. D. Lorenz, "Sensing Power MOSFET Junction Temperature Using Gate Drive Turn-On Current Transient Properties," *IEEE Transactions on Industry Applications*, vol. 52, no. 2, pp. 1677–1687, 2016.

[89] N. Baker, S. Munk-Nielsen, F. Iannuzzo, and M. Liserre, "Online Junction Temperature Measurement Using Peak Gate Current," in *2015 IEEE Applied Power Electronics Conference and Exposition (APEC)*, pp. 1270–1275, 2015.

[90] F. Yang, S. Pu, C. Xu, and B. Akin, "Turn-on Delay Based Real-Time Junction Temperature Measurement for SiC MOSFETs With Aging Compensation," *IEEE Transactions on Power Electronics*, vol. 36, no. 2, pp. 1280–1294, 2021.

[91] H. Chen, V. Pickert, D. Atkinson, and L. Pritchard, "On-Line Monitoring of The MOSFET Device Junction Temperature by Computation of the Threshold Voltage," in *The 3rd IET International Conference on Power Electronics, Machines and Drives, 2006. PEMD 2006*, pp. 440–444, 2006.

[92] B. Shi, S. Feng, L. Shi, D. Shi, Y. Zhang, and H. Zhu, "Junction Temperature Measurement Method for Power mosfets Using Turn-On Delay of Impulse Signal," *IEEE Transactions on Power Electronics*, vol. 33, no. 6, pp. 5274–5282, 2018.

[93] J. O. Gonzalez, O. Alatise, J. Hu, L. Ran, and P. A. Mawby, "An Investigation of Temperature-Sensitive Electrical Parameters for SiC Power MOSFETs," *IEEE Transactions on Power Electronics*, vol. 32, no. 10, pp. 7954–7966, 2017.

[94] J. O. Gonzalez, O. Alatise, J. Hu, L. Ran, and P. Mawby, "Temperature Sensitive Electrical Parameters for Condition Monitoring in SiC Power MOSFETs," in *8th IET International Conference on Power Electronics, Machines and Drives (PEMD 2016)*, pp. 1–6, 2016.

[95] S. Jahdi, O. Alatise, P. Alexakis, L. Ran, and P. Mawby, "The Impact of Temperature and Switching Rate on the Dynamic Characteristics of Silicon Carbide Schottky Barrier Diodes and MOSFETs," *IEEE Transactions on Industrial Electronics*, vol. 62, no. 1, pp. 163–171, 2015.

[96] Z. Zhang, F. Wang, D. J. Costinett, L. M. Tolbert, B. J. Blalock, and X. Wu, "Online Junction Temperature Monitoring Using Turn-Off Delay Time for Silicon Carbide Power Devices," in *2016 IEEE Energy Conversion Congress and Exposition (ECCE)*, pp. 1–7, 2016.

[97] Z. Zhang, J. Dyer, X. Wu, F. Wang, D. Costinett, L. M. Tolbert, and B. J. Blalock, "Online Junction Temperature Monitoring Using Intelligent Gate Drive for SiC Power Devices," *IEEE Transactions on Power Electronics*, vol. 34, no. 8, pp. 7922–7932, 2019.

[98] S. L. Rumyantsev, M. S. Shur, M. E. Levinshtein, P. A. Ivanov, J. W. Palmour, A. K. Agarwal, B. A. Hull, and S.-H. Ryu, "Channel Mobility and On-Resistance of Vertical Double Implanted 4H-SiC MOSFETs at Elevated Temperatures," *Semiconductor Science and Technology*, vol. 24, no. 7, p. 075011, Jun. 2009.

[99] Z. J. Wang, F. Yang, S. Campbell, and M. Chinthavali, "Development of a Low-Inductance SiC Trench MOSFET Power Module for High-Frequency Application," in *2018 IEEE Applied Power Electronics Conference and Exposition (APEC)*, pp. 2834–2841, 2018.

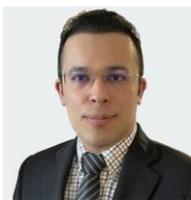

**Masoud Farhadi** (S'20) received the B.Sc. degree in electrical engineering with honors and the M.Sc. degree in power engineering (power electronics and systems) with honors from the Department of Electrical Engineering, University of Tabriz, Tabriz, Iran, in 2013 and 2016, respectively. He is currently working toward the Ph.D. degree at the University of Texas at Dallas, Richardson, TX, USA.

His current research interests include analysis and control of power electronic converters, reliability of power electronic systems, wide bandgap semiconductor device reliability, and renewable energy conversion systems.

Mr. Farhadi received the 2020 Jonsson School Industrial Advisory Council Fellowship, 2021 OK Kyun Kim and Youngmoo Cho Kim Graduate Fellowship and 2022 Excellence in Education Doctoral Fellowship from the University of Texas at Dallas.

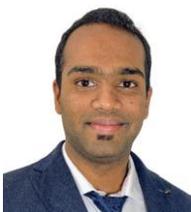

**Bhanu Teja Vankayalapati** (S'17) received the B.Tech degree in electrical engineering and the M.Tech degree in power electronics from the Indian Institute of Technology (BHU), Varanasi, in 2017, and the Ph.D. degree from The University of Texas at Dallas, Richardson, TX, USA, in 2022, in electrical engineering. From 2017 to 2018, he was a Project Engineer with the Indian Institute of Technology, Kanpur, India.

He is currently an Application Engineer with Texas Instruments Incorporated, Dallas, TX, USA. His research areas include WBG device applications, WBG reliability, converter design, and embedded control.

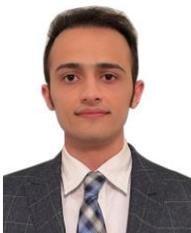

**Rahman Sajadi** received the B.Sc. degree in electrical engineering from Bu-Ali Sina University, Hamedan, Iran, in 2014, and the M.Sc. degree in electrical engineering from the University of Tehran, Tehran, Iran, in 2016. He is currently a Ph.D. student in electrical engineering with University of Texas at Dallas, Richardson, TX, USA. His research interests include design, modeling, reliability and control of power converters, wide bandgap semiconductor reliability, multilevel converters, and renewable energy systems.

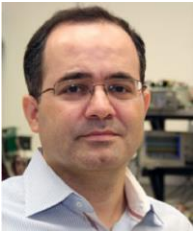

**Bilal Akin** (S'03-M'08-SM'13-F'22) received the Ph.D. degree in electrical engineering from the Texas AM University, College Station, TX, USA, in 2007. From 2005 to 2008, he was an RD Engineer with Toshiba Industrial Division, Houston, TX, USA. From 2008 to 2012, he was an RD Engineer with C2000 DSP Systems, Texas Instruments Incorporated. Since 2012, he has been with The University of Texas at Dallas, Richardson, TX, USA, as a Faculty. His research interests include design, control and diagnosis of electric motors and drives, digital power control and management, and fault diagnosis and condition monitoring of power electronics components and ac motors.

Dr. Akin is a recipient of NSF CAREER 2015 Award, IEEE IAS Transactions First Place Prize Paper Award and Top Editors Recognition Award from IEEE TVT Society, and two Jonsson School Faculty Research Awards. He is currently Area Editor of IEEE Transactions on Vehicular Technology.